\documentclass[fleqn]{article}
\begin{document}
\title{\bf Energy and angular momentum of the weak gravitational waves
 on the Schwarzschild background -- quasilocal gauge-invariant
formulation} 
\author{Jacek Jezierski\thanks{e-mail: Jacek.Jezierski@fuw.edu.pl}\\
Department of Mathematical Methods in Physics, \\ University of Warsaw,
ul. Ho\.za 74, 00-682 Warsaw, Poland
}
\maketitle


{\catcode `\@=11 \global\let\AddToReset=\@addtoreset}
\AddToReset{equation}{section}

\renewcommand{\theequation}{\thesection.\arabic{equation}}

\newtheorem{Lemma}{Lemma}
\AddToReset{Lemma}{section}
\newtheorem{Theorem}{Theorem}
\AddToReset{Theorem}{section}

\newfont{\msa}{msam10 scaled\magstep1}
\font\SYM=msbm10 
\def\Complex{\hbox{\SYM C}}
\def\Rationals{\hbox{\SYM Q}}
\def\Reals{\hbox{\SYM R}}
\def\Integers{\hbox{\SYM Z}}
\def\Naturals{\hbox{\SYM N}}

\newcommand{\Rho}{{\rm R}}
\newcommand{\dtwo}{\kolo{\Delta}}
\newcommand{\kolo}[1]{\vphantom{#1}\stackrel{\circ}{#1}\!\vphantom{#1}}
\newcommand{\ve}{\varepsilon}
\newcommand{\ol}{\overline}
\newcommand{\ind}{\mathop {\rm ind}\nolimits }
\newcommand{\be}{\begin{equation}}
\newcommand{\ee}{\end{equation}}
\newcommand{\ssr}[1]{{\, \strut ^{w}}\! {#1}}
\newcommand{\ssd}[1]{{\, \strut ^{d}}\! {#1}}
\newcommand{\ssm}[1]{{\, \strut ^{m}}\! {#1}}
\newcommand{\rd}{{\rm d}} 
\newcommand{\E}[1]{{\rm e}^{#1}}
\newcommand{\paragraf}[1]{\section{#1} \typeout{Paragraf \space
\thesection} }
\newcommand{\ten}[3]{{#1}^{#2}_{\ #3}}
\newcommand{\tenud}[3]{{#1}^{#2}{_{#3}}}
\newcommand{\tendu}[3]{{#1}_{#2}{^{#3}}}
\newcommand{\tenudu}[4]{{#1}^{#2}{_{#3}}{^{#4}}}
\newcommand{\tendud}[4]{{#1}_{#2}{^{#3}}{_{#4}}}
\newcommand{\wek}[2]{{#1}^{#2}}
\newcommand{\kowek}[2]{{#1}_{#2}}
\newcommand{\base}[2]{{{\partial}\over {\partial
{#1}^{#2}}}}
\newcommand{\diff}[2]{\frac{\partial{#1}}{\partial{#2}}}
\begin{abstract}
It is shown that the axial and polar perturbations of the
spherically symmetric black hole can be described in a
gauge-invariant way.  The reduced phase space describing
gravitational waves outside of the horizon is described by the
gauge-invariant quantities. Both degrees of freedom fulfill
generalized scalar wave equation. For the axial degree of freedom the
radial part of the equation corresponds to the Regge-Wheeler result
\cite{RW} and for the polar one we get Zerilli result
\cite{Zerilli}, see also  \cite{Chandra}, \cite{Moncrief} for both.
An important ingredient of the analysis is the {\em concept of
quasilocality} which does duty for the separation of the angular variables
in the usual approach. Moreover, there
is no need to represent perturbations by normal modes (with time
dependence $\exp(-ikt)$), we have fields in spacetime and the
Cauchy problem for them is well defined outside of the horizon.
The reduced symplectic structure explains the origin of the axial and
polar invariants. It allows to introduce an energy and angular
momentum for the gravitational waves which is invariant with respect to
the gauge transformations. Both generators
 represent quadratic approximation of the ADM nonlinear formulae
in terms of the perturbations of the Schwarzschild metric.
We also discuss the boundary-initial value problem for the linearized
Einstein equations on a Schwarzschild background outside of the horizon.
\end{abstract}

\section{Introduction}
In section 2 we introduce standard notions for linearized gravity and its
3+1 formulation. In the next section the analysis of constraints and
Killing fields on an initial surface gives charges for the linear field on
the Schwarzschild background. 

Section 4 contains the main technical results related to the so-called 2+1
decomposition of the initial data and to the gauge invariant
description of the evolution.
The invariants introduced in this section contain the full
gauge-independent information about initial data. If we ``insert'' the
initial value constraints into the canonical symplectic structure we can
express the symplectic 2-form $\Omega$ in terms of the invariants. This is
precisely shown in Appendix B. This way our invariants play a role of the
reduced 
unconstrained initial data which is gauge independent. Moreover, the axial
degree of freedom fulfills four-dimensional counterpart of the radial
equation proposed by Regge and Wheeler \cite{RW}, \cite{Chandra},
\cite{Moncrief} and the 
polar invariant is related to the Zerilli equation \cite{Zerilli},
\cite{Chandra}, \cite{Moncrief}. 

 Section 5 is devoted to the stationary
solutions, their behaviour on the horizon and precise interpretation of
the ``mono-dipole'' solutions. In
particular, the result of Vishveshwara \cite{Vish} that stationary
axial perturbation can exist only for dipole perturbation can be
easily recovered. However, if we assume more general conditions on a
horizon we may have other solutions.
 The dipole part of the axial
invariant corresponds to the spherical symmetry of the background
metric and represents angular momentum. On the other hand, there is
no dipole invariants in the polar part because the background metric is
not invariant with respect to the boosts and spatial
translations\footnote{Dipole invariants exist in flat Minkowski space and  
they represent linear momentum  and  center of mass respectively
 (see \cite{JJspin2}).}. Moreover,
the lack of invariant in the dipole polar part corresponds to the fact
that this part of the metric can be always ``gauged away''. We would
like to stress that mono-dipole perturbations of the Schwarzschild
metric represent different phenomena than the higher-pole
perturbations. It is clear from the approach that $l \geq 2$
represents {\em gravitational wave perturbation} whereas $l=0$ and
$l=1$ correspond to the {\em charges}. For example, $l=0$ in polar
part represents infinitesimal perturbation of the mass (we move in
the space of solutions from one Schwarzschild solution to another),
dipole part in axial part corresponds to the infinitesimal angular
momentum (we move from Schwarzschild to Kerr solution). Formally
also monopole part in axial degree of freedom can be analyzed and represents
Taub-NUT charge but this move is excluded by topology.

 Reduction of the symplectic
form presented in Appendix B allows to introduce invariants from the
symplectic point of view in section 6. In the next section we define
(in a gauge-invariant way)
hamiltonian system\footnote{After separation of the angular variables
several ``hamiltonian'' results proposed by the author can be easily
translated into Moncrief's approach presented in \cite{Moncrief}.}, energy
and angular momentum generators and 
boundary-initial value problem for the linearized field on the
Schwarzschild background. 
Moreover, we are able to show that the obtained hamiltonian is a quadratic
approximation of the ADM mass defined at spatial infinity for the full
nonlinear Einstein theory and this particular result will be described
elsewhere. 

The energy and angular momentum generators are well defined
for regular radiation data 
$\underline{\bf x}$, $\underline{\bf X}$, $\underline{\bf y}$,
$\underline{\bf Y}$ which is finite on the horizon and vanishing at spatial
infinity according to the S.A.F.
 Christodolou-Kleinerman condition \cite{Ch-Kl}.

In Appendix A we show how to reconstruct the full four-metric $h_{\mu\nu}$
from the invariants assuming the gauge conditions used in \cite{RW} and
\cite{Vish}. This construction explains the relations between our
invariants and the special form of the metric used in \cite{RW} and
\cite{Vish}. 
 In particular, we examine precisely ``mono-dipole'' part of
the metric ($l=0$ and $l=1$) which seems to be not fully analyzed in
literature.

\section{Linearized gravity}
In this section we remind some standard notions related to the Einstein
equations and the initial value problem.

Linearized
Einstein theory (see e.g. \cite{MTW} or \cite{Landau}) can be formulated as
follows.
Einstein equation
\be\label{nrE}
 2{\rm G}_{\mu \nu} (g) = 16\pi {\rm T}_{\mu \nu}
 \ee
 after linearization gives
\begin{equation}  \label{lrE}
h_{\mu \alpha}{_{;\nu}}{^{;\alpha}} +
         h_{\nu \alpha}{_{;\mu}}{^{;\alpha}} -
h_{\mu \nu}{^{;\alpha}}{_{\alpha}} -
(\eta ^{\alpha \beta}h_{\alpha \beta}){_{;\mu \nu}} - \eta _{\mu \nu}
[ h_{ \alpha\beta}{^{;\alpha \beta}} -
h_{\alpha}{^{\alpha ; \beta}}{_{\beta}} ] = 16\pi T_{\mu \nu} \end{equation}
\noindent
where pseudoriemannian metric ${\rm g}_{\mu \nu} = \eta _{\mu \nu} +
h_{\mu \nu}$,
$\eta _{\mu \nu}$ is the background metric and
 ``$;$'' denotes four--dimensional covariant derivative with respect
to the metric $\eta_{\mu \nu}$.
Moreover, we assumed that  $\eta _{\mu \nu}$ is a
vacuum solution of Einstein equation (${\rm G}_{\mu\nu}(\eta)=0$).

The (3+1)-decomposition of (\ref{lrE}) gives 6
dynamical equations for the space--space components $h_{kl}$ of the
metric (latin indices run from 1 to 3) and 4 equations which do not
contain time derivatives of $h_{kl}$. It is possible to introduce this
decomposition straightforward. However, a natural way for the formulation of
the (3+1)-splitting for the equation (\ref{lrE}) is to linearize ADM
formulation 
of the initial value problem for the full nonlinear Einstein equation
(\ref{nrE}). 
We shall introduce canonical variables for the
linearized case. They appear in a natural way if
we start from the ADM formulation of the initial value problem for
Einstein equations \cite{ADM}.

Let $({\rm g}_{kl}, {\rm P}^{kl})$ be the Cauchy data for Einstein
equations on a 
three--dimensional space-like surface $\Sigma$. This means
that ${\rm g}_{kl}$ is a Riemannian metric on $\Sigma$ and ${\rm P}^{kl}$
is a symmetric 
tensor density, which we identify with the ADM momentum \cite{ADM},
i.e.
\[
{\rm P}^{kl} = \sqrt{\det {\rm g}_{mn}} ({\rm g}^{kl} {\rm Tr} K -  K^{kl})
\]
where $K_{kl}$ is the second fundamental form (external curvature) of the
imbedding of $\Sigma$ into a spacetime $M$. \\
The 12 functions $({\rm g}_{kl}, {\rm P}^{kl})$ must fulfill 4 Gauss--Codazzi
constraints
\begin{equation}
     {\rm P}_i{^l}{_{| l}} = 8\pi\sqrt{\det g_{mn}}\, {\rm T}_{i\mu}n^{\mu}
                \label{ww}
\end{equation}
\begin{equation}
(\det {\rm g}_{mn}){\rm R}- {\rm P}^{kl}{\rm P}_{kl}
 + \frac{1}{2} ({\rm P}^{kl}{g}_{kl})^2  =
 16\pi(\det {\rm g}_{mn}) {\rm T}_{\mu \nu}n^{\mu}n^{\nu}
 \label{ws}
\end{equation}
where ${\rm T}_{\mu\nu}$ is an energy momentum tensor of the matter,
by $\rm R$ we denote the (three--dimensional) scalar curvature of
${\rm g}_{kl}$, $n^\mu$ is a future timelike four--vector normal to the
hypersurface 
$\Sigma$ and the calculations have been made with respect to the three--metric
${\rm g}_{kl}$ ("$|$" denotes the covariant derivative, indices are raised and
lowered etc.).

The Einstein equations and the definition of the metric connection imply
the first order (in time) differential equations for ${\rm g}_{kl}$ and
${\rm P}^{kl}$
(see \cite{ADM} or \cite{MTW} p. 525) and contain the lapse function
$N$ and the shift vector $N^k$ as parameters

\be
\label{gdot}
\dot{\rm g}_{kl}=\frac{2N}{\sqrt{\rm g}}\left( {\rm P}_{kl} -\frac 12
{\rm g}_{kl} {\rm P} \right) +
N_{k|l} +N_{l|k}
\ee
where ${\rm g}:= \det {\rm g}_{mn}$ and ${\rm P}:= {\rm P}^{kl}{\rm g}_{kl}$
\[
\dot{\rm P}^{kl} = -N\sqrt{\rm g} {\rm R}^{kl}
 + \sqrt{\rm g} \left( N^{|kl} -{\rm g}^{kl}N^{|m}{_{|m}} \right) +\]
\[+\frac 12 N\sqrt{\rm g} {\rm g}^{kl}{\rm R}
  - \frac{2N}{\sqrt{\rm g}}\left( {\rm P}^{km}{\rm P}_{m}{^l}-
\frac{1}{2} {\rm P} {\rm P}^{kl} \right) + \left( {\rm P}^{kl} N^m
\right)_{|m} + \] 
\be\label{Pdot}
+ \frac{N}{2\sqrt{\rm g}}{\rm g}^{kl} \left( {\rm P}^{kl}{\rm P}_{kl}-
\frac{1}{2}{\rm P}^2 \right)
-N^k{_{|m}} {\rm P}^{ml}-N^l{_{|m}} {\rm P}^{mk}
 + 8\pi N\sqrt{\rm g} {\rm T}_{mn}{\rm g}^{km}{\rm g}^{ln}
\ee

Let us consider an initial value problem for the linearized Einstein
equations on Schwarzschild background $\eta_{\mu\nu}$:
\be \label{Sch}
\eta_{\mu\nu}\, {\rm d}x^\mu \, {\rm d}x^\nu\, =
-\left(1-\frac{2m}r \right) {\rm d}t^2 + \left(1-\frac{2m}r\right)^{-1}
{\rm d}r^2  +r^2 {\rm d}\theta^2 + r^2 \sin^2 \theta {\rm d}\varphi^2
\ee
together with the radial coordinates: $x^3=r$,
$x^1=\theta $, $x^2=\varphi$. Moreover, $t=x^0$ denotes the time
coordinate. We consider only the part of the Schwarzschild spacetime outside
of the horizon, $r \geq 2m$.

We use the following convention for indices: greek indices $\mu, \nu,
\ldots$ run from 0 to 3;
$k,l, \ldots$ are spatial coordinates and run from 1 to 3;
$A,B,\ldots$ are spherical
angles $(\theta, \varphi)$ on a two-dimensional sphere $S(r):= \{
r=x^3=\mbox{const} \}$ and run from 1 to 2.
Moreover, let $\eta_{AB}$ denote a two-dimensional metric on $S(r)$.

Let $\displaystyle v:=1-\frac{2m}r$.
There are the following non-vanishing Christoffel symbols for the metric
(\ref{Sch}):
\[  \Gamma^{3}{_{33}}= -\frac{m}{vr^2}  \, ; \;\;
 \Gamma^{3}{_{AB}}= -\frac vr  \eta_{AB} \, ; \;\;
 \Gamma^{A}{_{3B}}= \frac 1r \delta^A{_B} \, ; \;\;
 \Gamma^{3}{_{00}}= \frac{mv}{r^2}  \, ; \;\;
 \Gamma^{0}{_{30}}= \frac{m}{vr^2}  \, ; \;\;  \Gamma^{A}{_{BC}} \]
 \noindent
where $\delta^A{_B}$ is the Kronecker's symbol
 and $\displaystyle \Gamma^{A}{_{BC}}$ are the same as for a standard
unit sphere $S(1)$ (in usual spherical coordinates
$\Gamma^{\theta}{_{\phi\phi}} = -\sin\theta\cos\theta$
 and $\Gamma^{\phi}{_{\phi\theta}}=\cot\theta$).

The curvature of the background metric we denote by
$C^\mu{_{\nu\lambda\kappa}}$ and the following
components of the Riemann tensor are non-vanishing (up to the symmetries
of the indices)
\[ C^0{_{A0B}}=C^3{_{A3B}}=\frac m{r^3}\eta_{AB} \, ; \quad
  C^{AB}{_{CD}}=\frac{2m}{r^3}\left( \delta^A{_C}\delta^B{_D}-
  \delta^A{_D}\delta^B{_C}  \right) \]

  We can introduce the following submanifolds of the Schwarzschild
spacetime $M$:  
  \be \label{Hs}
   \Sigma_s:=\{ x\in M \; : \; x^0=s, x^3 \geq 2m \} =\bigcup_{r \in
[2m,\infty[} S_s(r)
  \quad \mbox{where} \; \; S_s(r):= \{ x \in \Sigma_s \; : \;
x^3=r \} \ee
and $\Sigma_s$ is a partial Cauchy surface outside of the horizon.

The surface $\Sigma$ carries  the induced
Riemannian metric $\eta_{kl}$:
\be \label{eta3}
 \eta_{kl}\rd y^k \rd y^l=
 \frac 1v \rd r^2 +r^2 (\rd \theta^2 +\sin^2\theta \rd \phi^2)
 \ee
Usually it is convenient to change the coordinate $r$ to $r_*$ which is a
solution of an ordinary differential equation (see \cite{Chandra}): 
 \be \label{r*}  \frac{\rd r_*}{\rd r}=\frac 1v \, ; \; \quad
  r_* :=r+2m\ln \left(\frac r{2m}-1\right) \ee
and moves horizon to $-\infty$.

  The ADM momentum
${\rm P}^{kl}$  for the metric (\ref{eta3}) on each slice
$\Sigma_s$ vanishes (${\rm P}^{kl}=0$).
The shift vector  is also trivial ($N^k=0$) and the lapse $N=\sqrt v$ is
vanishing on the horizon. 
Moreover, the Ricci tensor for the three-metric $\eta_{kl}$
 has the following components:
\[ {\rm R}_{kl}=\frac1N N_{|kl} \]
 and the scalar curvature ${\rm R}$ vanishes.

Let us define the linearized variations $(h_{kl},P^{kl} )$ of the full
nonlinear Cauchy data $({\rm g}_{kl}, {\rm P}^{kl} )$ around background data
($\eta_{kl}$, 0)
\be  h_{kl}:={\rm g}_{kl}-\eta_{kl}\, ,\quad
 P^{kl}:={\rm P}^{kl}  \ee

We should now rewrite equations (\ref{ww})--(\ref{Pdot}) in a linearized
form in terms of $(h_{kl},P^{kl} )$.
Let us denote $P:=\eta_{kl}P^{kl}$ and  $h:=\eta^{kl}h_{kl}$.
The vector constraint (\ref{ww}) can be linearized as follows
\begin{equation} \label{ww1}
  {\rm P}_i{^l}{_{| l}} \approx P_i{^l}{_{| l}}
\end{equation}
Let us stress that the symbol ``$|$'' has  different meanings on the
left-hand side and on the right-hand side of the above formula.
It denotes the covariant derivative with
respect to the full nonlinear metric ${\rm g}_{kl}$
when applied to the ADM momentum ${\rm P}^{kl}$,
but on the right-hand side it means the covariant derivative with respect to
the background metric $\eta_{kl}$.
The scalar constraint (\ref{ws}) after linearization takes the form
\begin{equation}   \label{ws1}
\sqrt{\rm g}{\rm R}-\frac 1{\sqrt{\rm g}}\left( {\rm P}^{kl}{\rm P}_{kl}
 - \frac{1}{2} ({\rm P}^{kl}{\rm g}_{kl})^2 \right) \approx
 \sqrt\eta \left[ \left( h^{kl}{_{|l}}- h^{|k} \right)_{|k} -
 h^{kl}\Rho_{kl} \right]
\end{equation}
where $\sqrt\eta :=\sqrt{\det \eta_{kl}}$.

The linearized constraints for the vacuum ($T_{\mu\nu}=0$) have the form 
\begin{equation} \label{wwl}
  P_l{^k}{_{| k}}  =0
\end{equation}
\begin{equation}   \label{wsl}
  \left( h^{kl}{_{|l}}- h^{|k} \right)_{|k} -h^{kl}\Rho_{kl}  =0
\end{equation}
The linearization of (\ref{gdot}) leads to the equation
\begin{eqnarray}
\dot{h}_{kl} & = & \frac{2N}{\sqrt\eta} \left( P_{kl} -\frac 12
\eta_{kl} P \right) + h_{0k|l} +h_{0l|k}
\label{hdot}
\end{eqnarray}
where $N:=\frac 1{\sqrt{-\eta^{00}}}=\sqrt v$,
 $N_k=\eta_{0k}=0$ are the lapse and shift
for the background. Let us denote the linearized lapse by $n:=\frac12
{\sqrt v}h^0_0$. The linearization of (\ref{Pdot}) takes the form
\be
\frac 1{\sqrt\eta}\dot{P}_{kl} = -n {\rm R}_{kl} -N\delta R{_{kl}}
  + n_{|kl} -N_m\delta\Gamma^m{_{kl}} 
   - \eta_{kl}\left( n^{|m}{_{|m}}
-\eta^{ij}\delta\Gamma^m{_{ij}}N_{|m} -h^{mn}N_{|mn}  \right)
\label{pidot} \ee
where
\[ \delta\Gamma^m{_{kl}} :=\frac12\left(
h^m{_{k|l}}+h^m{_{l|k}}-h_{ml}{^{|k}} \right) \]
is the linearized Christoffel symbol and similarly
\[ \delta R{_{kl}} :=\frac12\left(
h^m{_{k|lm}}+h^m{_{l|km}}-h_{kl|m}{^{m}} -h_{|kl} \right) \]
is the linearized Ricci tensor.

It is well known (see for example \cite{JJspin2}) that the linearized Einstein
equations (\ref{lrE}) are invariant with respect to the ``gauge''
transformation: 
\begin{equation}
        h_{\mu \nu} \rightarrow h_{\mu \nu} + \xi _{\mu ; \nu} +
\xi _{\nu ; \mu}   \label{gauge4}
\end{equation}
where $\xi _{\mu}$ is a covector field.
 The (3+1)-decomposition of
the gauge acts on the Cauchy data in the following way
\begin{eqnarray}
2v\Lambda ^{-1} P_{kl} & \rightarrow & 2v\Lambda ^{-1} P_{kl}
 +(v \xi^0_{|k})_{|l}+(v \xi^0_{|l})_{|k} - \eta_{kl} (v\xi{^{0|m}})_m
    \label{pixi} \\
 \label{hxi}
 h_{kl}& \rightarrow & h_{kl} + \xi_{l | k}+ \xi_{k | l}
\end{eqnarray}
where  $\Lambda:={\sqrt v}\sqrt{\det \eta_{kl}}$ ($=r^2 \sin\theta$).
It can be easily checked that the scalar constraint (\ref{wsl}) and the vector
constraint (\ref{wwl}) are invariant with respect to the gauge
transformations (\ref{pixi}) and (\ref{hxi}).
The Cauchy data ($h_{kl}$, $P^{kl}$) and ($\overline{h}_{kl}$,
$\overline{P}^{kl}$) on $\Sigma$ are equivalent to each other
 if they can be related by the gauge transformation $\xi_{\mu}$.
 The evolution of canonical variables $P^{kl}$ and $h_{kl}$ given by
equations  (\ref{hdot}), (\ref{pidot})
 is not unique unless the lapse function $n$ and the shift vector
 $h_{0k}$ are specified.

We will show in the sequel that it is possible to define a reduced dynamics
in terms of invariants, which is no longer sensitive on gauge conditions.
The construction is analogous to the analysis given in \cite{GRG}.

\section{``Charges'' on the Schwarzschild background}

The vector constraint (\ref{wwl}) allows to introduce ``charges'' related
to the symmetries of the background metric. There are three generators of
the rotation group, which are simultaneously Killing vectors on the
initial surface $\Sigma$.
Let us denote this Killing field by $Z^k$. It is a solution of the Killing
equation
\be\label{Zk} Z_{k|l} + Z_{l|k} = 0 \ee
Let $V\subset \Sigma$ be a compact region in $\Sigma$. For example
$\displaystyle V:=\bigcup_{r\in [r_0,r_1]} S_s(r)$ and $\partial
V=S_s(r_0) \cup S_s(r_1)$.
 From (\ref{wwl}) and (\ref{Zk}) we get
\be\label{Lc}
0=\int_V P^{kl}{_{|l}} Z_k= \int_V (P^{kl}Z_k){_{|l}} =
 \int_{\partial V} P^{3k}Z_k
\ee
The equation (\ref{Lc}) expresses the ``Gauss'' law for the angular
momentum charge ``measured'' by the flux integral.
 It is easy to relate this charge to
the dipole part of invariant $\bf y$, which will appear in the sequel
 -- (\ref{y}) in the next section. For example, when $Z=\partial
/\partial\phi$ we have
\[
16\pi s^z:= 16\pi j^{xy} = -2\int_{\partial V} P^3{_\phi } =
 -2\int_{\partial V} P^3{_A} (r^2\varepsilon^{AB}\cos\theta)_{||B} =\]
\begin{equation} \label{sz}
 = 2\int_{\partial V} r^2 P^{3}{_{A||B}}\varepsilon^{AB} \cos\theta
 =\hspace{0.5cm} \int_{\partial V} \Lambda {\bf y} \cos\theta
 \end{equation}

The time translation defines a  mass charge from the scalar constraint
(\ref{wsl}) as follows
 \[
 0=\int_V
 N \sqrt\eta \left[\left( h^{kl}{_{|l}}- h^{|k} \right)_{|k}
-h^{kl}N_{|kl}\right] =
 \]
 \[
  =\int_V
  \left[
  N\sqrt\eta \left( h^{lk}{_{|k}}- h^{|l} \right) +\sqrt\eta
  \left(  N^{|l} h -N_{|k} h^{kl}  \right) \right]_{|l}  = \]
   \begin{equation}   \label{BM}
 = \int_{\partial V}
   \Lambda \left( h^{3k}{_{|k}}-  h^{|3} + \frac1N h N^{|3} - \frac1N
h^{k3}N_{|k}  \right)
\end{equation}
and it can be related to the monopole part of an invariant $\bf x$
introduced in the next section by (\ref{x}).

\[
16\pi p^0=\int_{\partial V}
   \Lambda \left( N h^{3k}{_{|k}}- N h^{|3}  + N_k h^{k3} -N^3 h
\right)= \] \begin{equation} \label{p0}
=\int_{\partial V} \frac{\Lambda}{r} \left(
 2 h^{33} -vr H,{_3} - (1-\frac{3m}r )H  \right) = 
  \hspace*{0.5cm} \int_{\partial V} \frac{\Lambda}r {\cal B}^{-1}{\bf x}
\end{equation}

\section{Equations of motion for the invariants}

The radial foliation of the Cauchy surface $\Sigma$ related to the
spherical symmetry allows to perform (2+1)-decomposition of the initial data.
In this section we introduce reduced gauge invariant data on $\Sigma$ for
the gravitational 
field, similar to the invariants introduced in \cite{GRG}. For this purpose
we use a spherical foliation of $\Sigma$ (see formulae (\ref{Hs}) and
(\ref{eta3})).

Let ${\dtwo}$ denotes the two--dimensional Laplace--Beltrami operator on a
unit sphere $S(1)$. Moreover, $H:=\eta^{AB}h_{AB}$, $\chi_{AB}:= h_{AB}
-\frac 12 \eta_{AB} H$, $S:=\eta^{AB}P_{AB}$, $S_{AB}:= P_{AB}
-\frac 12 \eta_{AB} S$ according to the notation used in \cite{GRG}.

The spatial gauge (\ref{hxi}) splits in the following way
\begin{eqnarray}
 \label{hxi33}
 h_{33}& \rightarrow & h_{33} + \frac2{\sqrt v}({\sqrt v}\xi_{3})_{,3}
\\ \label{hxi3A}
 h_{3A}& \rightarrow & h_{3A} + \xi_{3,A}+ \xi_{A,3} -\frac2r \xi_A
 \\ \label{hxiAB}
 h_{AB}& \rightarrow & h_{AB} + \xi_{A|| B}+ \xi_{B|| A} +\frac2r
\eta_{AB} \xi^3
\end{eqnarray}
where by ``$||$'' we denote the covariant derivative with respect to the
two--metric $\eta_{AB}$ on $S(r)$.
Similarly, the temporal gauge (\ref{pixi}) can be splitted as follows
\begin{eqnarray}
\Lambda ^{-1} P^{3}{_3} & \rightarrow & \Lambda ^{-1} P^3{_3}
 -\xi^{0||A}{_A} -\frac2r \xi^{0,3}
 \label{pxi33}  \\
 \label{pxi3A}
\Lambda ^{-1} P_{3A} & \rightarrow & \Lambda ^{-1} P_{3A} +
\left( \xi^0_{,3}-\frac1r \xi^{0} +\frac{m}{vr^2}\xi^0\right)_{||A}
  \\
  \label{sxiAB}
\Lambda ^{-1} S_{AB} & \rightarrow & \Lambda ^{-1} S_{AB}
  + \xi^0_{||AB} -\frac12\eta_{AB}\xi^{0||C}{_C}  \\
  \label{sxi}
  \Lambda ^{-1} S & \rightarrow & \Lambda ^{-1} S
 -\frac2N(N\xi^{0,3})_{,3} -\frac2r \xi^{0,3}   -\xi^{0||C}{_C} 
 \end{eqnarray}

 It is also quite easy to rewrite the (2+1)-decomposition of (\ref{hdot})
\begin{equation}
\dot{h}_{33} = \Lambda^{-1}(P^3{_3}-S) + \frac 2{\sqrt v}
({\sqrt v} h_{03}),{_3} \label{h33}
\end{equation}
\begin{equation}
\dot{h}_{3A} =2v\Lambda^{-1}P_{3A} + h_{03}{_{||A}} + h_{0A},{_3}
-\frac 2r h_{0A}   \label{h3A}
\end{equation}
\begin{equation}
\dot{h}_{AB}
= 2v\Lambda^{-1}S_{AB}- \eta_{AB}\Lambda^{-1}P^{33} +h_{0A||B} + h_{0B||A} +
 2vr^{-1} \eta_{AB} h_{03}      \label{hAB}
\end{equation}

The dynamical equations (\ref{pidot}) take the following (2+1)-form:

\[
2\Lambda^{-1}\dot{P}_{33}= -\frac 1v h^0{_0}{^{||A}}_A -2r^{-1}h^0{_0},{_3} +
\frac 1v h_3^{3 ||A}{_A} + 2r^{-2}h^3{_3}  + \]
\begin{equation} +  (H,{_3}- 2h_3{^{A}}{_{||A}} -2r^{-1}h_3{^3}),{_3}
+2r^{-1}(H,{_3}-2h_3{^A}{_{||A}}-2r^{-1}h_3{^3})        \label{P33}
\end{equation}
\[
2\Lambda^{-1}\dot{P}_{3C}= \left[\frac 1{\sqrt v} ({\sqrt v} h^0{_0}),{_3}
-r^{-1}h^0{_0}\right]_{ ||C} -\frac m{vr^2} h^3_{3||C} +
\frac{1}{2} (H,{_3}- 2h_3{^A}{_{||A}} -2r^{-1}h^3{_3})_{ ||C} +\]
\begin{equation}
+h_{3C||A}{^{||A}} +\frac 1{r^2} h_{3C} -
 \chi^A{_{C||A}},{_3} \label{P3C}
\end{equation}
\[
2\Lambda^{-1}\dot{S}_{AB} = h^0{_{0||AB}} -\frac12 \eta_{AB}h^0{_0}^{||C}{_C}
+ h^3{_{3||AB}} -\frac12 \eta_{AB}h^3{_3}^{||C}{_C}+\]
\[ - (h^3{_{A||B}}+h^3{_{B||A}}-
\eta_{AB}h^{3C}{_{||C}}),{_3}+(v\chi^C{_B},{_3}\eta_{CA}),{_3} +
\]
\begin{equation}
 + \chi_{AB}{^{||C}}_{||C} -
\chi^C{_{A||BC}} - \chi^C{_{B||AC}} + \eta_{AB}\chi^{CD}{_{||CD}} +
\frac 2{r^2}\chi_{AB}  \label{SAB}
\end{equation}

\[
2\Lambda^{-1} \dot S= -\sqrt{v} (\sqrt{v} h^0_{0,3})_{,3}-\frac 1r h^0_{0,3}
-h_0^{0||A}{_A} + (h^3{_3} +H)^{||A}{_A} +\frac2{r^2}(h^3{_3} +H)
 -\frac{12m}{r^3} h^3{_3}+
\]
\begin{equation}
 +\frac{2m}{r^2}h^3{_{3,3}}
+  v(H,{_3}- 2h_3{^{A}}{_{||A}} -2r^{-1}h_3{^3}),{_3}
+4\frac{v}r (H,{_3}-2h_3{^A}{_{||A}}-2r^{-1}h_3{^3}) -2\chi^{CD}{_{||CD}}
  \label{Sdot}
\end{equation}

 The vector constraint (\ref{wwl}) splits in a similar way
 \begin{equation}
\frac 1{\sqrt v}({\sqrt v}P^3{_3}),{_3} +  P_3{^A}{_{||A}} - r^{-1}S = 0
\label{ww3}
\end{equation}
\begin{equation}
P^3{_A},{_3} + S_A{^B}{_{||B}}+\frac{1}{2} S_{||A} = 0
\label{wwA}
\end{equation}

And finally
the (2+1)-decomposition of the scalar constraint (\ref{wsl}) can be written
in the form
\[
 h ^{|l}{_l} - h^{kl}{_{|kl}} +h^{kl} {\rm R}_{kl}
 = \frac{\sqrt v}{r^3}\left[ {r^2}{\sqrt v}
 (rH,{_3}-2rh_{3A}{^{||A}}-2h_3{^3})\right],{_3} +
\]
\begin{equation}
 +h^3{_{3 ||A}}{^A}+2r^{-2}h^3{_3} -\frac{6m}{r^3}h^3{_3}
 +\frac{1}{2}H^{||C}{_C}+r^{-2}H
  - \chi^{AB}{_{||AB}} = 0   \label{wsk}
\end{equation}

Let us notice that we can split the dynamics into two separate parts
which we call axial and polar respectively.
The axial part of initial data consists of two momentum components
 $P^{3A||B}\varepsilon_{AB}$, $S^C{_{A||CB}}\varepsilon^{AB}$ and two metric
components $h_{3A||B}\varepsilon^{AB}$, $\chi^C{_{A||CB}}\varepsilon^{AB}$. The
only gauge freedom is contained 
 in $\xi_{A||B}\varepsilon^{AB}$ which acts on the metric components.
 The gauge invariants $P^{3A||B}\varepsilon_{AB}$ and
 $S^C{_{A||CB}}\varepsilon^{AB}$ are related by the curl part of the vector
constraint  (\ref{wwA}) as follows:
\begin{equation}
(r^2P^{3A||B}\varepsilon_{AB}),{_3}
 + r^2 S^C{_{A||CB}}\varepsilon^{AB} =0
\label{wAB}
\end{equation}
It is easy to verify that
 the following pair of gauge invariants:
 \begin{equation}
 {\bf y}:=2\Lambda^{-1}r^2 P^{3A||B}\varepsilon_{AB} \label{y}
\end{equation}
\begin{equation}
 {\bf Y}:= \Lambda(\dtwo +2)h_{3A||B}\varepsilon^{AB}
 - r^2(\Lambda \chi^C{_{A||CB}}\varepsilon^{AB}),{_3} \label{Y}
\end{equation}
contains the whole information about axial part of initial data up to the
gauge freedom (see also Appendix A).
We will show in the sequel that this is a canonical pair with respect to the
symplectic structure of linearized Einstein equations.

Let us define
the polar invariants as follows
\begin{equation}
 {\bf x} := r^2\chi^{AB}{_{||AB}}-\frac{1}{2}(\dtwo+2) H+
  {\cal B}\left[ 2 h^{33}+2r h^{3C}{_{||C}} -rv H,{_3}  \right]
                \label{x}
\end{equation}
\begin{equation}
 {\bf X} :=
 2r^2 S^{AB}{_{||AB}} +{\cal B}\left[ 2rP^{3A}{_{||A}}
+\dtwo P^3{_3} \right]  \label{X}
\end{equation}
where
\[ {\cal B}:= (\dtwo +2)\left(\dtwo +2-\frac{6m}r\right)^{-1} \]
is a {\em quasilocal} operator -- it is local with respect to the
coordinates $t,r$ but non-local on each sphere $S(r)$.
 The proof that ${\bf x}$ and $\bf X$ contain
the whole information about polar part of initial data 
$P^{3A}{_{||A}}$, $S^{AB}{_{||AB}}$, $P^3{_3}$, $S$, 
 $h^{3C}{_{||C}}$, $\chi^{AB}{_{||AB}}$, $h^3{_3}$, $H$ up to the gauge
transformation $\xi^0, \xi^3, \xi^A{_{||A}}$ we give in Appendix A.
Moreover, in Appendix B we show that $({\bf x}, {\bf X}, {\bf y}, {\bf Y})$
is the reduced canonical initial data on $\Sigma$.

We can check the reduced field equations for the axial invariants

 \be\label{py} \dot{\bf y}=\frac{v}\Lambda {\bf Y}  \ee
\be\label{pY} \dot{\bf Y}= \Lambda \left\{
\left[\frac v{r^2}(r^2{\bf y}),{_3}\right],{_3} +\frac1{r^2}(\dtwo +2)
{\bf y}\right\} \ee
More precisely, the curl part of (\ref{P3C}) gives (\ref{py}) 
 and (\ref{pY}) follows directly from (\ref{h3A}), (\ref{hAB}) and may be
checked by inspection.

 It can be easily verified that the axial invariant $\bf y$ fulfills
the generalized scalar wave equation (as a consequence of the above dynamical
equations)
 \be \left(
 \mbox{\msa\symbol{3}} +\frac{8m}{r^3}\right) {\bf y}=0 \label{rfy} \ee
where $\mbox{\msa\symbol{3}}$ denotes the usual wave operator with respect to
the background metric $\eta_{\mu\nu}$.

There exists a simple relation between the equation (\ref{rfy}) and the
so-called {\em Regge-Wheeler equation} \cite{RW}, \cite{Chandra},
\cite{Moncrief}. 
Let us rewrite equation (\ref{rfy}) in the following form:
\be\label{RWE}
 -\ddot{\bf y} +\frac vr\left[v(r{\bf y})_{,3}\right]_{,3} = V^{(-)}
{\bf y} \ee
where $V^{(-)}$ is a ``spherical operator'' defined as follows:
\[ V^{(-)}:=-\frac v{r^2} \left(\dtwo+\frac{6m}r\right) \]
If we insert the invariant ${\bf y}$ in a special form
 ${\bf y} =\exp(i\sigma t) Y_l(\theta,\phi) Z^{(-)}(r)/r$
into equation (\ref{RWE}), we obtain Regge-Wheeler equation
\[ \left(\frac{d^2}{dr_*^2}+\sigma^2 \right)Z^{(-)}=V^{(-)}Z^{(-)} \]
Here $Y_l$ is a spherical harmonics such that
$\dtwo Y_l =-l(l+1) Y_l$ and coordinate $r_*$ is defined by (\ref{r*}).

The polar invariants fulfill the following equations:
 \be\label{px} \dot{\bf x}=\frac{v}\Lambda {\bf X}  \ee
\be\label{pX} \dot{\bf X}= \frac\Lambda{r^2} \left\{
\left( v{r^2}{\bf x},{_3}\right),{_3} + \left[\dtwo +v(1-2{\cal B}) +1 \right]
{\cal B}{\bf x}\right\} \ee
To obtain (\ref{px}) we need equations (\ref{h33}), (\ref{h3A}), (\ref{hAB})
and vector constraints (\ref{ww3}), (\ref{wwA}). Similarly, (\ref{pX}) is a
consequence of (\ref{P33}-\ref{SAB}) and scalar
constraint (\ref{wsk}) (see also Appendix A).

There exists also a generalized scalar wave equation for the polar
invariant but it is no longer local, it is only quasilocal:
 \be \left[\mbox{\msa\symbol{3}} +\frac{8m}{r^3}(\dtwo -1)
 \left(\dtwo +2-\frac{3m}r\right)\left(
 \dtwo +2-\frac{6m}r\right)^{-2}\right] {\bf x}=0
\label{rfx} \ee 

The similar equation to the (\ref{RWE}) can be presented in the analogous form
\be\label{ZE}
-\ddot{\bf x} +\frac vr\left[v(r{\bf x})_{,3}\right]_{,3} = V^{(+)}{\bf x}
\ee  
but now the operator $V^{(+)}$ is defined ``quasilocally''
\[ V^{(+)}:=-\frac {v}{r^2}  \left[(\dtwo+2)^2\left(\dtwo
-\frac{6m}r\right)+\frac{36m^2}{r^2}\left(\dtwo +2 -\frac{2m}r\right)\right] 
\left(\dtwo+2-\frac{6m}r\right)^{-2} \]
If we insert the invariant ${\bf x}$ in an analogous special form
 ${\bf x} =\exp(i\sigma t) Y_l(\theta,\phi) Z^{(+)}(r)/r$
into equation (\ref{ZE}) we obtain {\em Zerilli equation} \cite{Zerilli},
\cite{Chandra}, \cite{Moncrief} 
\[ \left(\frac{d^2}{dr_*^2}+\sigma^2 \right)Z^{(+)}=V^{(+)}Z^{(+)} \]
This way we have shown that both equations (Regge-Wheeler and Zerilli)
 posses gauge invariant formulations
and their solutions contain the entire gauge--independent information
about the linearized gravitational field on the Schwarzschild background
(see also \cite{Moncrief}).

Let us notice that $\bf x$ and $\bf y$ are scalars on each sphere $S_t(r)$
with respect to the coordinates $x^A$.
For the scalar $f$ on a sphere we can define a ``monopole'' part mon$(f)$
and a ``dipole'' part dip$(f)$ as a corresponding component with respect
to the spherical harmonics on $S^2$. Similarly, the ``dipole'' part of a
vector $v^A$ corresponds to the dipole harmonics for the scalars $v^A{_{||A}}$
and $\varepsilon^{AB}v_{A||B}$. Let us denote by $\underline f$ the
``mono--dipole--free'' part of $f$. According to this decomposition we have

\[ {\bf x}=\mbox{mon}({\bf x}) +\underline{\bf x}
\]
\[ {\bf y}= \mbox{dip}({\bf y}) +\underline{\bf y}
\]
The dipole part of $\bf x$ and monopole part of $\bf y$ are vanishing\footnote{
This is included in the definitions (\ref{x}) and (\ref{y}). More precisely,
$\bf y$ is a divergence (see also (\ref{dv})) and ${\bf x}=
r^2\chi^{AB}{_{||AB}}+ (\dtwo +2)[...]$. Moreover,
 dip$(\chi^{AB}{_{||AB}})=0$ because
double-divergence of any traceless tensor is ``mono-dipole-free'' (see also
(\ref{ddtt}) in Appendix B).}. 
The rest of the mono-dipole part of each scalar can be solved explicitly from
the equations (\ref{py})--(\ref{pX}) and the solution has the form
\[ {\bf x} - \underline{\bf x}= \frac{4{\bf m}}{r-3m}
\]
\[ {\bf y}-\underline{\bf y}=\frac{12{\bf s}}{r^2}
\]
From (\ref{py}), (\ref{px}) and the observation that
$\mbox{mon}({\bf X})=\mbox{dip}({\bf Y})=0$ we obtain
\[ \dot{\bf m}=\dot{\bf s}=0 \] 
Moreover, ${\dtwo m}=0$, $(\dtwo +2) {\bf s}=0$, which simply means that
$\bf m$ is a monopole and  $\bf s$ is a dipole, and they are constant with
respect to the coordinates $t,r$. They correspond to the charges
introduced in section 3. More precisely, ${\bf m}=p^0$.
Moreover, the angular momentum charge (\ref{sz}) can be obtained from
 the relation between spatial constant three-vector in
cartesian coordinates and dipole harmonics
\[  {\bf s} =\frac{s^l z_l}{r} \]
 where $z_l$ are cartesian coordinates, $r=\sqrt{\delta^{kl}z_k z_l}$
  and  $s^l$ is a corresponding three-vector representing angular momentum
(see \cite{JJspin2}). 

\section{Stationary solutions}
For the axial degree of freedom $\bf y$ we can rewrite equation
 ({\ref{RWE}) using a new coordinate $z:=\frac{2m}r$.
\[ 4m^2\ddot{\bf y}=(1-z)^2 z^4 \frac{\partial^2 {\bf y}}{\partial z^2}
 -(1-z) z^4 \frac{\partial {\bf y}}{\partial z} +
 (1-z) z^2(\dtwo +4z) {\bf y} \]
 This way horizon
corresponds to $z=1$ and spatial infinity to $z=0$. Let us consider
stationary solutions of the above equation which are regular
 at the spatial infinity (corresponding to $z=0$).
 \be\label{sty}
  (1-z) z^2 \frac{\partial^2 {\bf y}}{\partial z^2}
 - z^2 \frac{\partial {\bf y}}{\partial z} +
 (\dtwo+4z) {\bf y}=0 \ee
  Moreover, if we separate the
angular variables and include standard asymptotic behaviour at $z=0$:
\[ {\bf y} =z^{l+1} Y_l(x^A) u(z) \]
then the equation for the function $u$ is relatively simple
\[ (1-z) z u'' +[2l+1-(2l+3)z]u' -(l-1)(l+3)u =0 \]
and the solution regular at $z=0$ is given by the hypergeometric function 
\[ u=F(l-1,l+3,2l+2;z) \]
In particular for $l=1$ the function $u$ is constant, it represents
 the angular momentum charge solution (\ref{sz}), and 
corresponding $\bf y$ is finite on the horizon. On the other hand,
for $l\geq 2$ we obtain logarithmic divergence of the hypergeometric
function $F$ at $z=1$. More precisely,
 \[ F(l-1,l+3,2l+2;z) = z^{-2l-1}[P(z) + \tilde P (z)\ln(1-z)] \]
 where $P$ and $\tilde P$ are polynomials. The solution is not regular at
$z=1$ and it confirms the result of
Vishveshwara \cite{Vish} that the only nontrivial stationary perturbation
is given by the axial perturbation with $l=1$. The interpretation of
this solution is given by (\ref{sz}).
 One can check by direct
computation for the Kerr metric \cite{Kerr}, \cite{B-L}:
\[  \rho ^2 := r^2 + a^2 \cos^2 \theta \quad \triangle := r^2 -2mr +a^2 \]
\[ g_{00} = -1 +{2mr \over \rho ^2}  \quad g_{0\phi} = -{2mra\sin ^2 \theta
\over \rho ^2} \quad g_{rr} = {\rho ^2 \over \triangle} \quad
g_{\theta \theta} = \rho^2 \]
\[ g_{\phi\phi} =
\sin^2\theta\left(r^2+a^2+{2mra^2\sin^2\theta\over\rho^2}\right) \] 
 that the infinitesimal
angular momentum gives the same result. More precisely, 
 the linear part of the Kerr metric $g_{\mu\nu}$
 with respect to the parameter $a$:
\[ g_{\mu\nu}\rd x^\mu\rd x^\nu =
 \eta_{\mu\nu}\rd x^\mu\rd x^\nu - 4\frac{ma}r\sin^2\theta \rd
t\rd\phi + O(a^2) \]
gives only the one non-vanishing component of the linearized Kerr metric on the
Schwarzschild background:
 \[ h_{0\phi}=-\frac{2ma}r \sin^2\theta \]
From the definition (\ref{y}) and the equation (\ref{h3A}) we can
calculate the invariant ${\bf y}=12ma\cos\theta/r^2$.
It is easy to compare the result with (\ref{sz}) and it gives
angular momentum charge $s^z=ma$.

Can we consider monopole solution in axial invariant?
The definition (\ref{y}) in terms of the initial data does not allow
nontrivial monopole part of the axial invariant.
However, we can consider such situation if we admit singular metric.
Formally this case corresponds to the infinitesimal Taub-NUT solution. The
Taub-Nut metric \cite{Taub}, \cite{NUT}:

\[ g_{\mu\nu}\rd x^\mu\rd x^\nu =
(r^2 +l^2)(\rd\theta^2 +\sin^2\theta\rd\phi^2) +\tilde v^{-1} \rd r^2 -
\tilde v (\rd t +2l\cos\theta\rd\phi)^2 = \] \[
= \eta_{\mu\nu}\rd x^\mu\rd x^\nu
-4vl\cos\theta \rd t \rd\phi + O(l^2) \, ;\quad
 \tilde v := {r^2 -2mr -l^2 \over r^2 + l^2} \]
gives the linearized metric $h_{0\phi}=-2lv\cos\theta$ and the monopole
axial invariant
$\displaystyle {\bf y}=-\frac{4l}r(1-\frac{3m}r)$. We should stress that
this is only 
formal calculation because the tensor $h_{0\phi}=-2lv\cos\theta$ is not
well defined along the $z-$axis and is excluded as a global solution.
The monopole charge in $\bf y$ plays a role of the topological 
obstruction for the existence of the global metric. It is similar to the
magnetic monopole in electrodynamics (see also \cite{JJspin2}).

For polar degree of freedom $\bf x$ from equation (\ref{ZE}) we
get
$$ 4m^2\ddot{\bf x}=(1-z)^2 z^4 \frac{\partial^2 {\bf x}}{\partial z^2}
 -(1-z) z^4 \frac{\partial {\bf x}}{\partial z} +
 (1-z) z^2\left[\frac13+\frac23(\dtwo-1)(\dtwo+2)(\dtwo+2-3z)^{-2}
\right](\dtwo +2) {\bf x} $$
and the corresponding stationary equation has the following form
\be\label{stx}
 (1-z) z^2 \frac{\partial^2 {\bf x}}{\partial z^2}
 - z^2 \frac{\partial {\bf x}}{\partial z} +
 \left[\frac13+\frac23(\dtwo-1)(\dtwo+2)(\dtwo+2-3z)^{-2}
\right](\dtwo +2) {\bf x}=0 \ee
For $l=0$ the solution $\displaystyle {\bf x}=\frac{z}{2-3z}$ is related to
(\ref{p0}) and corresponds to the mass charge. More precisely, if we put
in the metric (\ref{Sch}) 
$m+\delta m$ instead of parameter $m$ and take the linear part in $\delta
m$ we get
\[ \eta_{\mu\nu}(m+\delta m)\rd x^\mu\rd x^\nu =
 \eta_{\mu\nu}(m)\rd x^\mu\rd x^\nu + \frac{2\delta m}r \rd t^2 + 
 \frac{2\delta m}{v^2r} \rd r^2 + O(\delta m^2) \]
 and the invariant $\displaystyle {\bf x}= 2{\cal B} h^{33}=\frac{4\delta m}r
\left(1-\frac{3m}r\right)^{-1}$.
 If we compare with mass charge (\ref{p0}) we obtain
$p^0=\delta m$ (see also \cite{RW}).

For $l\geq 2$ we have the following transformation law:
\[ \dtwo(\dtwo+2){\bf x}=-6z^2(1-z)\frac{\partial {\bf y}}{\partial z} +
\left[ \dtwo(\dtwo+2) +6z(1-z)-18z^2(1-z)(\dtwo+2-3z)^{-1} \right] {\bf y}
\]
which moves the solution of (\ref{sty}) into solutions of (\ref{stx}).
In particular, it is clear that for $l\geq 2$ the stationary solutions of
the equation (\ref{stx}) have the same logarithmic divergence on the
horizon as the solutions of the equation (\ref{sty}).
The explicit stationary solutions (in a specific gauge) were also given by
Zerilli in \cite{Zerilli} and here we present a gauge invariant confirmation
of his result.

\underline{Remark} Although
 $l=1$ is excluded in the definition of $\bf x$,
 we can consider another variable 
 \[ (\dtwo +2)^{-1}{\bf x}:= (\dtwo +2)^{-1}r^2\chi^{AB}{_{||AB}} -\frac
12 H +\left(\dtwo +2-\frac{6m}r\right)^{-1}
[2h^{33} +2rh^{3A}{_{||A}}-rvH_{,3}] \]
which is no longer gauge-invariant in its dipole part
(see also (\ref{trg}) in Appendix A). More precisely, it transforms with
respect to the gauge transformation as follows
\[ \mbox{dip}\left(\frac12 H +\frac{r}{6m}vQ\right)=
 \mbox{dip}(-(\dtwo +2)^{-1}{\bf x}) \longrightarrow \mbox{dip}(-(\dtwo
+2)^{-1}{\bf x}) + \mbox{dip}(\xi^A{_{||A}}) \]
where $Q$ is defined in Appendix A by (\ref{defQ}).
Formally, the dipole solution $(\dtwo +2)^{-1}{\bf x}=\ln(1-z) Y_1(x^A)$
  fulfills the same equation (\ref{stx}), and it 
corresponds to the Regge-Wheeler infinitesimal translation as a gauge
transformation\footnote{If we assume that the translation corresponds to
the $\xi^3=\cos\theta$ then the component $\xi^\theta$ is uniquely defined
as the polar gauge transformation preserving 
the gauge condition $h_3{^A}_{||A}=0$ which has been used by Regge-Wheeler
 \cite{RW} and Vishveshwara \cite{Vish}. }
 (eq. 32 in \cite{RW}):
\[ \xi^3=\cos\theta \, ; \quad \xi^{\theta}=\frac{\sin\theta}{2m}\ln
\left(1-\frac{2m}r\right) \, ; \quad \xi^A{_{||A}}=\frac{\cos\theta}m \ln 
\left(1-\frac{2m}r\right)
\]
This way the stationary polar dipole solution corresponds to the
infinitesimal translation gauge and it is also logarithmically divergent
on the horizon.

\section{The symplectic form and its reduction}

In this section we show the relation between the symplectic structure and the
invariants introduced in the section 4.
 Let $(P^{kl},h_{kl})$ be the  Cauchy data on a hypersurface $\Sigma$.
  Let us consider the symplectic structure
  \[ \Omega:= \int_{\Sigma}\delta{P}^{kl}\wedge \delta{h}_{kl} \]  
 It is invariant with respect to the spatial gauge transformation which
 is fixed on the boundary ($\left.\delta\xi^k\right|_{\partial {\Sigma}}=0$):
\begin{eqnarray}
 \int_{{\Sigma}}\delta{P}^{kl}\wedge\delta{h}_{kl} 
 & \stackrel{\xi^k}{\longrightarrow} &
  \int_{{\Sigma}} \delta{P}^{kl}\wedge\delta{h}_{kl} + 
  2 \int_{\partial {\Sigma}}\delta{P}^3{_l}\wedge \delta{\xi}^{l} \nonumber
 \end{eqnarray}
Moreover, it is invariant with respect to the temporal gauge if we assume
that $\xi^0$ and its normal derivative\footnote{You may ``correct'' the
symplectic form by a boundary term in such a way that the result 
is gauge invariant for $\xi^{\mu}$ vanishing on the $\partial {\Sigma}$ without an
extra assumption about derivatives (see \cite{Kij-qlh}).}
  are fixed on the boundary:
  \begin{eqnarray}
 \int_{{\Sigma}}\delta{P}^{kl}\wedge\delta{h}_{kl} & 
 \stackrel{\xi^0}{\longrightarrow} &
  \int_{{\Sigma}}\delta{P}^{kl}\wedge\delta{h}_{kl} +\nonumber \\
  & &   +\int_{\partial {\Sigma}}\sqrt{\eta}\left[
  \delta (N\xi^0)_{|k}\wedge\delta h^{3k} -\delta(N\xi^0)^{|3}\wedge\delta h 
  + N\delta\xi^0\wedge \delta(h^{|3}-h^{3l}{_{|l}}) \right] \nonumber
  \end{eqnarray}
Roughly speaking, the symplectic structure is invariant with
respect to the gauge modulo boundary terms.
 The quadratic form $\int_{{\Sigma}}\delta{P}^{kl}\wedge\delta{h}_{kl}$
can be decomposed into monopole part, dipole part and the remainder in a
natural way.

 From the considerations given in the Appendix B (formulae (\ref{rap}) and
(\ref{rpp})) we can easily see that the ``radiation'' part
\[ \underline{\Omega}=
\int_{{\Sigma}}\delta\underline{P}^{kl}\wedge\delta\underline{h}_{kl} \sim
\int_{{\Sigma}}\delta\underline{\bf X}\wedge 
\dtwo ^{-1}(\dtwo +2)^{-1}\delta\underline{\bf x}+
 \delta\underline{\bf Y}\wedge
 \dtwo ^{-1}(\dtwo +2)^{-1} \delta\underline{\bf y} \]
where symbol ``$\sim$'' denotes equality modulo 
boundary term. Moreover, the ``mono-dipole'' part has the form
(see (\ref{mpp}) and (\ref{dap}) respectively)
 \be\label{mfs}
 \mbox{mon}( \int_{{\Sigma}}\delta{P}^{kl}\wedge\delta{h}_{kl}) =  
 \int_{{\Sigma}}
\frac1{2}\delta{P}_{33}\wedge \delta {\cal B}^{-1}\mbox{mon}({\bf x} )+
\frac12 \int_{\partial {\Sigma}} r \delta P^3{_3}\wedge\delta \mbox{mon}(H)
\ee

 \be\label{dfs}
 \mbox{dip}(\int_{{\Sigma}}\delta{P}^{kl}\wedge\delta{h}_{kl}) \sim
 -\int_{{\Sigma}} \Lambda \delta\mbox{dip}({\bf y}) \wedge
 \dtwo ^{-1}\delta({h}_{3A||B}\varepsilon^{AB})
\ee

The monopole part of the polar invariant $\mbox{mon}({\bf x} )$ and
the dipole part of the axial one
 $\mbox{dip}({\bf y})$ represent mass and angular momentum respectively, and
 they are supposed to be fixed. These quantities are analogous to the
electric charge in electrodynamics. If we assume that there is no matter
inside volume ${\Sigma}$ 
then both of them are fixed by the constraints, provided that they are
controlled at $r=2m$ as a boundary condition on the horizon. Let us assume
that $\left. \delta\mbox{mon}({\bf x} )\right|_{S(r=2m)}=0$, 
$\left. \delta \mbox{dip}({\bf y})\right|_{S(r=2m)}=0$ then
 the mono-dipole part (\ref{mfs}) and (\ref{dfs})
vanish and the symplectic
structure reduces to the ``mono-dipole-free'' invariants
\be\label{ssonS}
 \int_{\Sigma}\delta{P}^{kl}\wedge\delta{h}_{kl} \sim \int_{\Sigma}
\delta\underline{\bf X}\wedge\dtwo ^{-1}(\dtwo +2)^{-1}\delta\underline{\bf x}+
\delta\underline{\bf Y}\wedge\dtwo ^{-1}(\dtwo +2)^{-1} \delta\underline{\bf y}
\ee
This way we obtain $\underline{\bf X}, \underline{\bf x},
 \underline{\bf Y}, \underline{\bf y}$ as the quasi-local canonical
variables describing reduced unconstrained initial data on $\Sigma$.

\underline{Remark} One can ask the question when (\ref{ssonS}) is a strict
equality not only modulo boundary term.
 The symplectic 2-form $\Omega$
reduces to the mono-dipole-free invariants if we assume the following
boundary conditions which fix the gauge freedom on the boundary:
 \[ \left. \delta h_{AB}\right|_{\partial {\Sigma}}=0 \quad
 \left. \delta(2h^3_3+2rh_3{^A}{_{||A}}-rH_{,3})\right|_{\partial{\Sigma}}=0 
 \]
 \be\label{bd}
  \left. \delta(P^{3A||B}\varepsilon_{AB}) \right|_{\partial {\Sigma}}=0 \quad
   \left. \delta (P^3_3 + 2r{\dtwo}^{-1}P^{3A}{_{||A}} )
    \right|_{\partial {\Sigma}}=0  \ee
 In Appendix A we analyze the possibility when $\underline H$, 
 $\underline Q$, $\underline \Pi$ and
$\chi^{AC}{_{||CB}} \varepsilon_A{^B}$ are precisely the gauge conditions
and then $\chi^{AB}{_{||AB}}$ corresponds to the control of
$\underline{\bf x}$. 
Roughly speaking, the ``control mode'' given by (\ref{bd}) contains
four boundary conditions related to the gauge freedom plus two boundary
conditions for the unconstrained degrees of freedom which we propose to call
Dirichlet boundary data.

If we introduce quasilocal coordinates:
\[ q^1:=\dtwo ^{-1/2}(\dtwo +2)^{-1/2}\underline{\bf x} \, ; \quad
 p_1:=\dtwo ^{-1/2}(\dtwo +2)^{-1/2}\underline{\bf X} \]
\[ q^2:=\dtwo ^{-1/2}(\dtwo +2)^{-1/2}\underline{\bf y} \, ; \quad
 p_2:=\dtwo ^{-1/2}(\dtwo +2)^{-1/2}\underline{\bf Y} \]
we can rewrite the reduced symplectic structure (\ref{ssonS})
 in the canonical form  
\[ \Omega = \int_{\Sigma}
\delta\underline{\bf X}\wedge \dtwo ^{-1}(\dtwo
+2)^{-1}\delta\underline{\bf x}+ 
 \delta\underline{\bf Y}\wedge \dtwo ^{-1}(\dtwo +2)^{-1}
\delta\underline{\bf y} = \sum_{n=1}^2
\int_{\Sigma}  \delta p_n \wedge \delta q^n
 \]
 Unconstrained initial data for the full nonlinear theory (which is
similar to the considerations in this article) has been proposed in
\cite{CQG94}. Moreover, the concept of quasilocality appeared
in \cite{Kyoto91} and has been developed in \cite{Kij-qlh}.  The boundary
data possesses its counterpart in the full nonlinear theory (see
proposition in \cite{Kij-qlh}) and it will be discussed elsewhere.

\section{Energy and angular momentum of the gravitational waves}

The reduction of the symplectic form (\ref{ssonS}) allows to
formulate the hamiltonian relation in terms of the reduced canonical
variables 
 \[  \int_{\Sigma} \dot{\bf x}\dtwo ^{-1}(\dtwo +2)^{-1}  \delta {\bf X}
 - \dot{\bf X} \dtwo ^{-1}(\dtwo +2)^{-1} \delta {\bf x} + \]
 \[  + \int_{\Sigma} \dot{\bf y}\dtwo ^{-1}(\dtwo +2)^{-1}  \delta {\bf Y}
 - \dot{\bf Y} \dtwo ^{-1}(\dtwo +2)^{-1} \delta {\bf y} =  16\pi\delta
{\cal H} + \] 
 \be  + \int_{\partial {\Sigma}} \frac{\Lambda}r v (r{\bf x}) ,_{3}
 \dtwo ^{-1}(\dtwo +2)^{-1} \delta {\bf x}
 +\frac{\Lambda}{r} v (r{\bf y}) ,_{3}
 \dtwo ^{-1}(\dtwo +2)^{-1} \delta {\bf y} \ee
where
\[ 16\pi{\cal H}:= \frac12 \int_{\Sigma} \frac v\Lambda {\bf X}
 \dtwo ^{-1}(\dtwo +2)^{-1} {\bf X} + \frac v\Lambda {\bf Y}
 \dtwo ^{-1}(\dtwo +2)^{-1} {\bf Y}+
\]
\[  + \frac1{2} \int_{\Sigma} \frac{\Lambda}{r^2}\left[ v (r{\bf x})_{,3}
 \dtwo ^{-1}(\dtwo +2)^{-1} (r{\bf x})_{,3} + 
  {\bf x} \frac{r^2}v V^{(+)}\dtwo ^{-1} (\dtwo +2)^{-1}{\bf x} \right] +\]
\be + \frac1{2} \int_{\Sigma} \frac\Lambda{r^2} \left[ v (r{\bf y})_{,3}
 \dtwo ^{-1}(\dtwo +2)^{-1} (r{\bf y})_{,3} + {\bf y}
\frac{r^2}v V^{(-)}\dtwo ^{-1} (\dtwo +2)^{-1} {\bf y} \right]
 \label{Hxy} \ee
 (see also eq. 4.19 and 5.34 in \cite{Moncrief}). \\
Similarly for angular momentum we propose the following expression
\be\label{JZ} 16\pi J(Z)=\int_\Sigma {\bf X} \dtwo ^{-1}(\dtwo +2)^{-1}
 Z^A\partial_A{\bf x} + {\bf Y} \dtwo ^{-1}(\dtwo +2)^{-1}
 Z^A\partial_A{\bf y}
\ee
where $Z$ is a Killing field (\ref{Zk}). In particular for $Z=\partial
/\partial\phi$ the $z-$component of the angular momentum takes the form:
\be\label{Jxy} 16\pi J_z=\int_\Sigma {\bf X} \dtwo ^{-1}(\dtwo +2)^{-1}
{\bf x}_{ ,\phi} + {\bf Y} \dtwo ^{-1}(\dtwo +2)^{-1}
{\bf y}_{ ,\phi}
\ee
The conservation laws for the energy and angular momentum
\[ \partial_0 {\cal H} = 0
\]
\[ \partial_0 J(Z) = 0
\]
are fulfilled if we assume appropriate boundary conditions on the horizon
and at the spatial infinity. The natural choice of those conditions is to
assume that $\underline{\bf x}$ and $\underline{\bf y}$ are vanishing on
the boundary $\partial\Sigma$.

After separation of the angular variables in (\ref{Hxy})
 we obtain a hamiltonian which has
been used for the energy method by Wald \cite{Wald} (see also in
\cite{Chandra} eq. 386) and it confirms stability of the Schwarzschild
metric (see also \cite{Moncrief}).
Here we have shown how to combine an energy of different multipoles together.

\subsection{Regular initial data} 
If we assume that the invariants ${\bf x}$, ${\bf y}$ are vanishing on
the horizon then we get a nice hamiltonian system outside of the horizon.
We can also assume that $\underline{\bf x}$ and $\underline{\bf y}$ are
fixed and finite on the horizon. More precisely, we propose the following
initial-boundary data:\\
1. mon($\bf x$)=0, because of the singularity at $r=3m$,
moreover this charge indicates that we have chosen wrong parameter $m$ in
the background.\\
2. dip($\bf y$) -- weak internal angular momentum.\\
 3. The radiation data $(\underline{\bf x}$, $\underline{\bf X}$,
$\underline{\bf y}$, $\underline{\bf Y})$
has to be finite on the horizon, moreover $\underline{\bf x}$,
$\underline{\bf y}$  should be controlled as a Dirichlet boundary condition.
We assume also that $h_{\mu\nu}=O(1/r)$ at spatial infinity.
The center of mass contained in dip$[(\dtwo+2)^{-1}{\bf x}]$ and
linear momentum in dip$(\Pi)$ can be always ``gauged out'', see Appendix A.1,
and those gauge transformations correspond to the infinitesimal
translation and boost respectively. Performing those gauge transformations
we pass to the ``center mass rest frame''.  \\
4. The asymptotics of the invariants $\underline{\bf x}$, $\underline{\bf y}$
at spatial infinity should guarantee finite hamiltonian and this can be
achieved for standard asymptotics $h_{\mu\nu}=O(1/r)$. 
Unfortunately, the standard asymptotics $\frac 1r$ does not guarantee
finite angular momentum. More precisely, standard asymptotics at spatial
infinity $\underline{\bf x}, \underline{\bf y}=O(1/r)$,
$\underline{\dot{\bf x}}, \underline{\dot{\bf y}}=O(1/r^2)$
gives logarithmically divergent integral in (\ref{Jxy}). 
The proposition by Christodolou and Kleinerman \cite{Ch-Kl}, so-called
S.A.F. condition, fits perfectly if we adopt it to the linearized theory.
We propose the following asymptotics at spatial infinity
\[ {\bf x},{\bf y}=O(r^{-3/2})\, , \quad \dot{\bf x},\dot{\bf y}
=O(r^{-5/2}) \]  
\underline{Remark}
Christodolou and Kleinerman assume that the full ADM data has better
asymptotics (except conformal factor in the riemannian metric). Let us
notice that if we assume that $P^{kl}=O(r^{-1/2})$ and
$h_{kl}=O(r^{-3/2})$ then all boundary 
terms in the symplectic form (analyzed in Appendix B) are vanishing like
$\frac 1r$ at spatial infinity.

\section{Conclusions}
We have shown a natural functionals which represent energy and angular
momentum of the weak gravitational radiation on a Schwarzschild background.
Particularly, we do not have to separate angular variables and the result is
gauge-invariant.
Moreover, the equations of motion for the gauge-invariant degrees of
freedom correspond to the well-known Regge-Wheeler and Zerilli results. We
give also a complete (together with the interpretation) analysis of the
stationary solutions.

After separation of the angular variables we should notice that the
results presented in sections 4, 6 and 7 are very close to the approach
presented by Moncrief in \cite{Moncrief}. However, the invariants used by
Moncrief in polar part are different and non-local.

\appendix
\section{Reduction of the initial data to the invariants}

It is convenient to introduce the following variables:
\be\label{defPi} \Pi:=2rP^{3A}{_{||A}} +\dtwo P^3{_3} \ee
\be\label{defQ} Q:= 2h^3_3 +2rh_3{^A}{_{||A}}-rH_{,3} \ee
The scalar constraint (\ref{wsk}) takes the form
\be\label{wskq} \frac{\sqrt v}r(r^2\sqrt{v} Q)_{,3} +r^2\chi^{AB}{_{||AB}}
-\frac 12 (\dtwo +2)H -\left(\dtwo+2-\frac{6m}r\right)h^3_3 =0 \ee
From vector constraint (\ref{ww3}-\ref{wwA}) we get
\be\label{wwPi}
 r{\sqrt v}(\sqrt{v}\Pi)_{,3}+\left(
 \dtwo+2-\frac{6m}r\right)rP^{3A}{_{||A}}+2vr^2
S^{AB}{_{||AB}} =0 \ee

\subsection{Dipole polar part of the initial data}
In this subsection we consider only dipole parts of the variables and we
denote by the same letter their dipole parts as the full objects in the
rest of the paper.

1. From vector constraint (\ref{wwPi}), (\ref{ww3}) and the definition
(\ref{defPi})  we have
\be\label{dP3A} rP^{3A}{_{||A}}=\frac{r^2\sqrt{v}}{6m}(\sqrt{v}\Pi)_{,3} \ee
\be\label{dP33}
P^3{_3}=\frac{r^2\sqrt{v}}{6m}(\sqrt{v}\Pi)_{,3}-\frac12 \Pi \ee
\be\label{dS} S=\left[
\frac{r^2\sqrt{v}}{6m}(\sqrt{v}\Pi)_{,3}\right]_{,3} \ee
This means that $\Pi$ contains the full information about dipole polar
part of $P^{kl}$. 

2. From scalar constraint (\ref{wskq}) we get
\[ h^3_3=-\frac{\sqrt v}{6m}(r^2\sqrt{v} Q)_{,3} \]
Moreover, from the definition (\ref{defQ}) we have
\[ 2rh_3{^A}{_{||A}}= Q+ \frac{\sqrt v}{3m}(r^2\sqrt{v} Q)_{,3} +rH_{,3} \]
This way we can see that $Q$ and $H$ contain the full information
about dipole polar part of the linearized metric $h_{kl}$.

3. The polar dipole gauge can be described by the following transformations:
\[ -\frac{r^3}{12m\Lambda}\Pi \longrightarrow -\frac{r^3}{12m\Lambda}\Pi
   + \xi^0 \]
\[ -\frac{rv}{6m} Q \longrightarrow  -\frac{rv}{6m} Q + \xi^3 \]
\be\label{trg} \frac 12 H +\frac{rv}{6m} Q \longrightarrow \frac 12 H
+\frac{rv}{6m} Q  +\xi^A{_{||A}} \ee
 and they show that we can always perform quasilocal gauge transformation
in such a way that $\Pi$, $Q$ and $H$ are vanishing.

Moreover, from (\ref{h33}--\ref{P3C}) we can check the evolution equations:
\[ -h^0_0 = \frac{r^3}{6m\Lambda}\dot\Pi +\frac 16 Q \]
\[ h_{03}=\frac{r^2 v}{12m\Lambda}(r\Pi)_{,3} -\frac{r^2}{12m}\dot Q \]
\[ h_0{^A}{_{||A}}=\frac 12 \dot{H} +\frac{rv}{6m} \dot{Q}
-\frac{rv}{6m\Lambda}\Pi \]
Finally we have shown that quasilocal gauge $\Pi=Q=H=0$ gives vanishing
dipole polar part of the full metric $h_{\mu\nu}$.

\subsection{Radiation polar part of the initial data in Regge-Wheeler gauge}

The special form of the metric $h_{\mu\nu}$ proposed in \cite{RW} will be
called Regge-Wheeler gauge\footnote{
Regge-Wheeler impose maximal number of gauge conditions:
 1 axial gauge (\ref{RWag}) and 3 polar gauge conditions (\ref{RWpg})
in radiation part. They do not discuss mono-dipole part except some stationary
solutions. Chandrasekhar in his book \cite{Chandra} does not impose axial
gauge. Vishveshwara uses the same formalism as Regge-Wheeler
plus axial gauge (\ref{Vag}) for the dipole and moreover in polar part
he assumes dip$(h_0^0+h^3_3)=0$. The monopole part is not discussed in the
literature.}. In polar part it gives the following gauge
conditions: 
\[
\chi^{AB}{_{||AB}}=h_{0A}{^{||A}}=h_{3A}{^{||A}}=0
\, ; \quad r^2\chi^{AB}{_{||AB}} \longrightarrow r^2\chi^{AB}{_{||AB}}
+(\dtwo +2)\xi^A{_{||A}} \]
\be\label{RWpg}
r^2 h_{0A}{^{||A}} \longrightarrow r^2 h_{0A}{^{||A}}
+\dtwo\xi_0 +r^2 \dot\xi^A{_{||A}} \, ; \quad
r^2 h_{3A}{^{||A}} \longrightarrow r^2 h_{3A}{^{||A}}
+\dtwo\xi_3 +r^2 (\xi^A{_{||A}})_{,3}
\ee
which allow to reconstruct polar part of the metric $h_{\mu\nu}$ as follows:

1. Equation
\[ 2v\Lambda^{-1}S^{AB}{_{||AB}}=\dot\chi^{AB}{_{||AB}} -
 (\dtwo +2)h_{0A}{^{||A}} \]
  follows directly from (\ref{hAB}) and
 gives  $S^{AB}{_{||AB}}=0$. The variable $\Pi$ (defined by (\ref{defPi}))
  contains the same information as
$\underline{\bf X}=2r^2S^{AB}{_{||AB}}+{\cal B}\underline\Pi$.
 From (\ref{wwPi}) we
reconstruct $\underline{P}^{3A}{_{||A}}$, and 
(\ref{defPi}) gives $\underline{P}^3{_3}$. Finally,  
we reconstruct $\underline S$ from vector
constraint (\ref{ww3}). This way we have obtained four polar components of
the ADM momentum $\underline{P}^{kl}$, namely $S^{AB}{_{||AB}}$, 
$\underline{P}^{3A}{_{||A}}$, $\underline{P}^3{_3}$ and $\underline S$.
The remaining two axial components are analyzed in subsection A.4.

2. The spatial metric we reconstruct from the scalar constraint (\ref{wskq})
together with (\ref{defQ}) and the observation that ${\cal B}Q={\bf x}
+\frac12(\dtwo +2)H$.  
More precisely, we get a system of two equations for two missing components
$\underline H$, $\underline h^3{_3}$ of the spatial metric (we need
boundary data to solve them!). 

3. The lapse $\underline h^0_0$ we get from $\dot{S}^{AB}{_{||AB}}$ which
is given by (\ref{SAB}). And finally the missing component $\underline
h_{03}$ of the shift we can calculate from the following equation
\[ 2r^2\Lambda^{-1}P^{3A}{_{||A}}+\dtwo h_{03}=r^2\dot{h}_{3A}{_{||A}}
 - r^2({h}_{0A}{^{||A}})_{,3} \] 
 which may be easily checked from (\ref{h3A}). The axial components of the
spatial metric and axial part of the shift vector we analyze in subsection
A.4.

The above analysis together with the results in subsection A.4 show
 how to reconstruct the full initial data 
 $(\underline{P}^{kl}, \underline{h}_{kl})$ together with lapse 
 $\underline{h}^0_0$ and shift $\underline{h}_{0k}$ in the Regge-Wheeler
gauge (\ref{RWpg}) and (\ref{RWag}) from the reduced initial data 
($\underline{\bf x}$, $\underline{\bf y}$, $\underline{\bf X}$,
$\underline{\bf Y}$).

\subsection{Radiation polar part of the initial data in quasilocal gauge}
We do not like to impose any conditions directly on lapse and shift.
It is more elegant to impose gauge conditions on the initial data only.
The ``time conservation laws'' of the gauge conditions obtained from
equations of motion give lapse and shift indirectly.
For this purpose we propose the following quasilocal gauge conditions
\[ \underline Q=\underline H=\underline\Pi=0 \]
which allow to reconstruct radiation polar part of the four-metric in a
quasilocal way.
\[ rvQ \longrightarrow rvQ +2\left(\dtwo+2-\frac{6m}r\right)\xi^3 \]
\[ \frac{r^2}\Lambda \Pi \longrightarrow 
 \frac{r^2}\Lambda \Pi -\frac12\dtwo\left(
 \dtwo+2-\frac{6m}r\right)\xi^0 \]
 \[ vQ-\frac12 \left(\dtwo+2-\frac{6m}r\right)H
  \longrightarrow vQ-\frac12 \left(\dtwo+2-\frac{6m}r\right)H
   -\left(\dtwo+2-\frac{6m}r\right)\xi^A{_{||A}} \]
 Moreover, each component of the metric $\underline{h}_{\mu\nu}$ depends
quasilocally on the invariants. This can be shown as follows:\\
1. From (\ref{defQ}) and (\ref{wskq}) we obtain metric components 
\[ r^2\chi^{AB}{_{||AB}}=\underline{\bf x} \, , \quad \underline{H}=0 \]
\[ \underline{h}^3_3=-r\underline{h}_{3A}{^{||A}} =\left(\dtwo +2
-\frac{6m}r\right)^{-1} \underline{\bf x} \] 
2. From (\ref{defPi}), (\ref{wwPi}) and (\ref{ww3}) we get the ADM momentum
\[ 2r^2 S^{AB}_{||AB}=\underline{\bf X} \, , \quad \underline S=\left[
2vr\dtwo ^{-1} \left(
\dtwo +2 -\frac{6m}r\right)^{-1}\underline{\bf X} \right]_{,3}
-\dtwo ^{-1}\underline{\bf X} 
 \]
\[ \dtwo \underline{P}^3_3=-2r\underline{P}^{3A}{_{||A}} =2v\left(\dtwo +2
-\frac{6m}r\right)^{-1} \underline{\bf X} \] 
3. Moreover,
\[ r\Pi -\Lambda r\dot Q= \left(\dtwo +2 -\frac{6m}r\right)(rP^3{_3} -2\Lambda
h_{03}) 
\] and 
\[ v\dot Q -\frac12 \left(\dtwo +2 -\frac{6m}r\right)\dot H =\Lambda^{-1} v\Pi
-\left(\dtwo +2 -\frac{6m}r\right) h_0{^A}{_{||A}}
\]
give the following components of the shift vector
\[ \underline{h}_{0A}{^{||A}}=0 \, , \quad
\underline{h}_{03}=\frac{v}{\Lambda}\dtwo ^{-1}\left(\dtwo +2
-\frac{6m}r\right)^{-1}\underline{\bf X}\] 
and finally the equation for $\dot\Pi$ gives the lapse function
\[ \underline{h}^0{_0}=\dtwo ^{-1}\left(\dtwo +2 -\frac{6m}r\right)^{-1} \left(
 \dtwo \underline{\bf x} - 
2v{\cal B}\underline{\bf x} -2rv\underline{\bf x}_{,3} \right) \]
The above quasilocal formulae for the radiation polar part of the metric
$h_{\mu\nu}$  allow to check the equations of motion (\ref{px}) and
(\ref{pX}) by inspection from the (\ref{h33}-\ref{SAB}).

Why we prefer the quasilocal gauge?\\
Because the initial data with compact support becomes more clear.
The full (constrained) initial data with compact support gives the reduced
initial data with compact support and the opposite is true only in
quasilocal gauge. The data with compact support allows to avoid the
boundary value problems.

\subsection{Monopole part of the initial data}
The monopole part of the data seems to be not analyzed in the literature.
We propose here the complete analysis of this simple ``gap''.
Let us prolongate the gauge condition $H=0$ which fixes the radial coordinate
and for the monopole part let us assume
\[ \mbox{mon}(H)=0 \]
This way $\mbox{mon}({\bf x})= 2v{\cal B}\mbox{mon}(h^3{_3})$ and 
the monopole part of the metric takes the form
\[\mbox{mon}( H)=0 \, ; \quad \mbox{mon}(h^{33})=\frac12{\cal B}^{-1}
\mbox{mon}({\bf x})=2\frac{p^0}r \] 

For time coordinate we propose the following gauge condition
\[ \mbox{mon}(P^3{_3})=0 \]
which is no longer quasilocal, it needs the boundary data at spatial
infinity for the parabolic equation obtained from (\ref{pxi33}).
The similar situation (nonlocal reconstruction)
 we encounter during analysis of the dipole axial part
 (see the next subsection). 

From the vector constraint (\ref{ww3}) and  $\mbox{mon}(P^3{_3})=0$
we obtain
\[ \mbox{mon}(S)=0 \]
This way the whole ADM momentum is trivial in its monopole part.

The trace of (\ref{hAB}):
\[ \frac12 \dot H =\Lambda^{-1} P^{33} +h_{0A}^{||A} +\frac{2v}r h_{03} \]
gives $\mbox{mon}(h_{03})=0$, and finally the lapse we get from 
monopole part of (\ref{P33})
\[\mbox{mon}( 2\frac{r^2}{\Lambda}\dot{P}^3{_3}) =
\mbox{mon}( -2rv h^0_{0,3} +2vh^{3}_3)  \]
If we assume that $h^0_0$ vanishes at spatial infinity we obtain
\[ \mbox{mon}(h_0^0)= \frac{p^0}{m}\ln v \]
The infinity which we encounter in the monopole part of the invariant 
${\bf x}$ at $r=3m$ suggests that we have to assume $p^0=0$ and this way
we exclude the possibility of manipulation with the mass parameter $m$ in the
background metric.

\subsection{Axial part of the initial data}
1. The ADM momentum components $P^{3A||B}\varepsilon_{AB}$ and
$S^{AC}{_{CB}}\varepsilon_A{^B}$ we get from $\bf y$ and axial part of the
vector constraint (\ref{wAB}).\\ 
2. In radiation part ($l\geq 1$) we impose  gauge condition
\be\label{RWag}
\chi^{AC}{_{||CB}}\varepsilon_A{^B}=0 \, ; \quad
r^2 \chi^{AC}{_{||CB}}\varepsilon_A{^B} \longrightarrow
r^2\chi^{AC}{_{||CB}}\varepsilon_A{^B} + (\dtwo +2)
\xi_{A||B} \varepsilon^{AB}
\ee
 which fixes $\xi_{A||B}\varepsilon^{AB}$ quasilocally.
 The component $h_{3A||B} \varepsilon^{AB}$ is obviously contained in $\bf
Y$. Moreover, from (\ref{hAB}) we have
\[ r^2 \dot\chi^{AC}{_{||CB}}\varepsilon_A{^B} =
2v\frac{r^2}\Lambda S^{AC}{_{||CB}}\varepsilon_A{^B} + (\dtwo +2)
 h_{0A||B} \varepsilon^{AB} \]
 which gives $h_{0A||B} \varepsilon^{AB}=0$.
 This way we have shown how to reconstruct axial part of the metric
$\underline{h}_{\mu\nu}$ from the invariants $(\underline{\bf y},
\underline{\bf Y})$. \\
 3. The dipole part of the metric ($l=1$) can be fixed by the gauge condition
\be\label{Vag}
\mbox{dip}(h_{3A||B}\varepsilon^{AB}) =0  \, ; \quad
\mbox{dip}(h_{3A||B}\varepsilon^{AB}) \longrightarrow
\mbox{dip}(h_{3A||B}\varepsilon^{AB}) +
 \left[ \mbox{dip}(\xi_{A||B}\varepsilon^{AB}) \right]_{,3}
\ee
 which gives ``parabolic'' equation for the angular
 gauge transformation $\mbox{dip}(\xi_{A||B}\varepsilon^{AB})$ with the
 boundary value at spatial infinity. Moreover, stationary solution in
dip$({\bf y})$ appears in $h_{0A||B}\varepsilon^{AB}$. More precisely, from
``time conservation law of gauge condition''
\[ 0=r^2\mbox{dip}(\dot{h}_{3A||B}\varepsilon^{AB})=
  r^2\mbox{dip}({h}_{0A||B}\varepsilon^{AB})_{,3}
  +\mbox{dip}({\bf y}) \]
we obtain dipole axial part of the shift
 $\displaystyle\mbox{dip}({h}_{0A||B}\varepsilon^{AB})=\frac{4{\bf s}}{r^3}$, 
and in particular for ${\bf s}=s\cos\theta$ we have
$\displaystyle h_{0\phi}=-\frac{2s\sin^2\theta}{r}$ which has been proposed by
Vishveshwara (eq. 5.2 in \cite{Vish}).

\section{Reduction of the symplectic form}
 Let $(p^{kl},q_{kl})$ denotes the Cauchy data on a hypersurface $\Sigma$.
The (2+1)-splitting of the tensor $q_{kl}$ gives the following
components on a sphere: $\stackrel{2}{q}:=\eta^{AB}q_{AB}$, $q_{33}$ --
scalars on $S^2$, $q_{3A}$ -- vector and
$\stackrel{\circ}{q}\! {_{AB}}:=
q_{AB}-\frac12\eta_{AB}\stackrel{2}{q}$ -- symmetric traceless tensor on
$S^2$. Similarly, we can decompose the tensor density $p^{kl}$.
On each sphere $S(r)$ we can manipulate as follows
 \[
  \int_{V}{p}^{kl}{q}_{kl} = \int_{V} {p}^{33}{q}_{33} +
 2{p}^{3A}{q}_{3A}
+ \frac{1}{2} \stackrel{2}{p} \stackrel{2}{q} +
\stackrel{\circ}{p}\! {^{AB}}\stackrel{\circ}{q}\! {_{AB}}  =
\]
\[
=  \int_{V} {p}^{33}{q}_{33} -
2(r{p}^{3A}{_{|| A}}) {\dtwo} ^{-1} (r{q}_{3A}{^{|| A}}) -
 2(r{p}^{3A|| B}\varepsilon_{AB}) {\dtwo} ^{-1}
(r{q}_{3A|| B}\varepsilon^{AB}) +
\frac{1}{2} \stackrel{2}{p} \stackrel{2}{q} +
\]
\[
+ 2 \int_{V} (r^2\varepsilon^{AC}\stackrel{\circ}{p}\! {_A{^B}}{_{|| BC}})
 {\dtwo} ^{-1}({\dtwo}+2) ^{-1}
(r^2\varepsilon^{AC}\stackrel{\circ}{q}\! {_A{^B}}{_{|| BC}}) + \]
\[
+ 2 \int_{V} (r^2\stackrel{\circ}{p}\! {^{AB}}{_{|| AB}})
 {\dtwo} ^{-1}({\dtwo}+2) ^{-1}
 (r^2\stackrel{\circ}{q}\! {^{AB}}{_{|| AB}})
 \]
 where we have used the following identities on a sphere
 \be\label{dv}
  -\int_{S(r)} \pi^A v_A = (r{\pi}^{A}{_{|| A}}) {\dtwo} ^{-1}
(rv^{A}{_{|| A}})+ (r{\pi}^{A|| B}\varepsilon_{AB}) {\dtwo} ^{-1}
(rv_{A|| B}\varepsilon^{AB}) 
 \ee
 and similarly for the traceless tensors we have
 \[ 
    \int_{S(r)} \stackrel{\circ}{\pi}\! ^{AB} \stackrel{\circ}{v}\! _{AB} =
  2 \int_{S(r)} (r^2\varepsilon^{AC}\stackrel{\circ}{\pi}\! {_A{^B}}{_{|| BC}})
 {\dtwo} ^{-1}({\dtwo}+2) ^{-1}
(r^2\varepsilon^{AC}\stackrel{\circ}{v}\! {_A{^B}}{_{|| BC}}) + \]
 \be\label{ddtt}
+ 2 \int_{S(r)} (r^2\stackrel{\circ}{\pi}\! {^{AB}}{_{|| AB}})
 {\dtwo} ^{-1}({\dtwo}+2) ^{-1}
 (r^2\stackrel{\circ}{v}\! {^{AB}}{_{|| AB}})
 \ee
 The axial part of the quadratic form $\int_{V}{p}^{kl}{q}_{kl}$
 we define as follows:
 \[ \mbox{axial part}= -2\int_{V}(r{p}^{3A|| B}\varepsilon_{AB})
{\dtwo} ^{-1} (r{q}_{3A|| B}\varepsilon^{AB}) +\]
 \[
+ 2 \int_{V} (r^2\varepsilon^{AC}\stackrel{\circ}{p}\! {_A{^B}}{_{|| BC}})
 {\dtwo} ^{-1}({\dtwo}+2) ^{-1}
(r^2\varepsilon^{AC}\stackrel{\circ}{q}\! {_A{^B}}{_{|| BC}})  \]
 The remainder we define as a polar part. Moreover, from the axial part of
the vector constraint (\ref{wAB}):
 \be
 (r^2 {p}^{3A|| B}\varepsilon_{AB}),{_3} +
 r^2 \varepsilon^{AC}\stackrel{\circ}{p}\! {_A{^B}}{_{|| BC}} =0 \ee
 we obtain
\[ \mbox{axial part}  = -2\int_{V} 
 (r{p}^{3A|| B}\varepsilon_{AB}) {\dtwo} ^{-1}
(r{q}_{3A|| B}\varepsilon^{AB}) + \]
\[
 - 2 \int_{V} (r^2{p}^{3A|| B}\varepsilon_{AB}),{_3}
 {\dtwo} ^{-1}({\dtwo}+2)^{-1}
(r^2\varepsilon^{AC}\stackrel{\circ}{q}\! {_A{^B}}{_{|| BC}}) =
\]
  \[ =
 -2 \int_{\partial V}  (r^2{p}^{3A|| B}\varepsilon_{AB})
 {\dtwo} ^{-1}({\dtwo}+2)^{-1}
(r^2\varepsilon^{AC}\stackrel{\circ}{q}\! {_A{^B}}{_{|| BC}}) + \]
  \[
  -2\int_{V} (r^2{p}^{3A|| B}\varepsilon_{AB})
{\dtwo} ^{-1}[{q}_{3A|| B}\varepsilon^{AB}  - 
   ({\dtwo}+2)^{-1} (r^2\varepsilon^{AC}\stackrel{\circ}{q}\!
{_A{^B}}{_{|| BC}}),{_3} ]  \]
We can see the invariants in the volume term if we write dipole part
separately 
\be\label{dap} \mbox{dipole axial part} =
 -2 \int_{V}\mbox{dip} (r^2{p}^{3A|| B}\varepsilon_{AB})\dtwo ^{-1}
({q}_{3A|| B}\varepsilon^{AB}) \ee
and finally the radiation axial part contains gauge-invariants in the
volume term
 \[
\mbox{radiation axial part} =
 -2 \int_{\partial V}  (r^2{p}^{3A|| B}\varepsilon_{AB})
 {\dtwo} ^{-1}({\dtwo}+2)^{-1}
(r^2\varepsilon^{AC}\stackrel{\circ}{q}\! {_A{^B}}{_{|| BC}}) + \]
  \be\label{rap}
  -2\int_{V} (r^2{p}^{3A|| B}\varepsilon_{AB})
{\dtwo} ^{-1}({\dtwo}+2)^{-1}[({\dtwo}+2){q}_{3A|| B}\varepsilon^{AB}  - 
    (r^2\varepsilon^{AC}\stackrel{\circ}{q}\!
{_A{^B}}{_{|| BC}}),{_3} ]  \ee
 
 For the polar part we can use the rest of the vector constraints
 \be\label{p333}
 \frac{r}{\sqrt v} ({\sqrt v}p^3{_3})_{,3} + r{p}_{3A}{^{|| A}} - 
\stackrel{2}{p} = 0 \ee
\be\label{p3AA}
(r^2 {p}^{3A}{_{|| A}}),{_3} + (r^2 \stackrel{\circ}{p}\!
{^{AB}}{_{|| AB}})  + \frac{1}{2}\dtwo  \stackrel{2}{p} = 0 \ee
 and we can reduce partially polar part as follows
 
 \[ \mbox{polar part} = \int_{V} {p}^3{_3}{q}_3{^3} -
2(r{p}^{3A}{_{|| A}}) {\dtwo}^{-1} (r{q}_{3A}{^{|| A}})  
+ \frac{1}{2}\left( \frac{r}{\sqrt v} ({\sqrt v}p^3{_3})_{,3} +
r{p}^{3A}{_{|| A}}\right) 
\stackrel{2}{q} +
\]
\[
-2 \int_{V} \left[(r^2{p}^{3A}{_{|| A}}),{_3} + \frac{1}{2}{\dtwo}
 \left( \frac{r}{\sqrt v} ({\sqrt v}p^3{_3})_{,3}
  + r{p}_{3A}{^{|| A}}\right) \right]
 {\dtwo}^{-1}({\dtwo}+2)^{-1}
 (r^2\stackrel{\circ}{q}\! {^{AB}}{_{|| AB}}) =
 \]
 \[
= \int_{\partial V} r{p}^3{_3}\left[ \frac{1}{2}\stackrel{2}{q} -
({\dtwo}+2)^{-1} (r^2\stackrel{\circ}{q}\! {^{AB}}{_{|| AB}}) \right]
-2 \int_{\partial V}  (r^2{p}^{3A}{_{|| A}}) {\dtwo}^{-1}({\dtwo}+2)^{-1}
 (r^2\stackrel{\circ}{q}\! {^{AB}}{_{|| AB}}) + \]
  \[
  + \int_{V} r{p}_{3A}{^{|| A}} \left[ \frac{1}{2}\stackrel{2}{q} +
 2r v{\dtwo} ^{-1} ({\dtwo}+2)^{-1} (r^2\stackrel{\circ}{q}\!
{^{AB}}{_{|| AB}}),{_3} -  ({\dtwo}+2)^{-1} (r^2\stackrel{\circ}{q}\!
{^{AB}}{_{|| AB}}) - 2  {\dtwo} ^{-1} (r{q}^{3A}{_{|| A}}) \right] +  \]
\be\label{pp}
  + \int_{V}{p}^3{_3} \left[ {q}_3{^3}
  + {\sqrt v}({\dtwo}+2)^{-1} \left(\frac{r^3}{\sqrt v}
\stackrel{\circ}{q}\! {^{AB}}{_{|| AB}}\right),{_3}  
  - \frac{\sqrt v}{2}\left( 
  \frac{r}{\sqrt v}\stackrel{2}{q}\right),{_3} \right] \ee
The above calculation shows that we should consider mono-dipole part
separately. The monopole part is very simple
 \be\label{mpp}
 \mbox{mon}( \int_{V}{p}^{kl}{q}_{kl})=  \int_{V}
 \frac12 {p}_{33} {\cal B}^{-1} \mbox{mon}(\zeta )
+\frac{1}{2} \int_{\partial V} r p_3{^3} \mbox{mon}(\stackrel{2}{q})
\ee
 where invariant $\zeta$ is defined as follows
 \[ \zeta:=  {\cal B}\left[ 2q^{33} + 2r {q}^{3A}{_{|| A}}
 -rv \stackrel{2}{q},{_3} \right]
 + r^2\stackrel{\circ}{q}\! {^{AB}}{_{|| AB}}
   -\frac12 (\dtwo +2) \stackrel{2}{q}  \]
 
 Using  (\ref{dP33}), (\ref{dS}), (\ref{dP3A}) and integrating by parts we
 can also rewrite (from the beginning) dipole part in the following way
 \[
 \mbox{dipole polar part} =  \int_{V} {p}^{33}{q}_{33} -
2(r{p}^{3A}{_{|| A}}) {\dtwo}^{-1} (r{q}^{3A}{_{|| A}}) 
 + \frac{1}{2} \stackrel{2}{p} \stackrel{2}{q} =\]
 \[ = \int_{V}\left[ \frac{r^2\sqrt{v}}{6m}(\sqrt{v}\Pi)_{,3}-\frac12\Pi
\right] {q}_{33} - 2 \frac{r^2\sqrt{v}}{6m}(\sqrt{v}\Pi)_{,3} {\dtwo}^{-1}
(r{q}^{3A}{_{|| A}})  
 + \frac{1}{2}\left[ \frac{r^3\sqrt{v}}{6m}(\sqrt{v}\Pi)_{,3} \right]_{,3}
\stackrel{2}{q} =\] 
\[ = \int_{\partial V} \frac{r^3\sqrt{v}}{12m}(\sqrt{v}\Pi)_{,3} 
\stackrel{2}{q} 
+ \frac{r^2}{6m}\Pi \left(r{q}^{3A}{_{|| A}}+q^{33} 
-\frac12 vr\stackrel{2}{q}_{,3}\right) + \]
\[ - \int_{V}\Pi\left\{ \frac12 q^3_3 +\frac{\sqrt
v}{6m}\left[r^2\sqrt{v}\left( q^3_3 -\frac12
r\stackrel{2}{q}_{,3}+rq_3{^A}{_{||A}}\right) \right]_{,3}\right\} = 
  \int_{\partial V} 
\frac{r^3\sqrt{v}}{12m}(\sqrt{v}\Pi)_{,3} \stackrel{2}{q} + 
\frac{r^2 v}{12m} \Pi Q =\]
\be\label{dpp}
= \int_{\partial V} r v \Pi\dtwo ^{-1}\left(\dtwo +2-\frac{6m}r\right)^{-1} Q
 -r^2 p^{3A}{_{||A}}\dtwo ^{-1} \stackrel{2}{q} 
 \ee
where here $\Pi$ denotes only its dipole part and $\Pi$ itself is defined
by (\ref{defPi}).
We have also used scalar constraint (\ref{wskq}), and finally the dipole
polar part takes its boundary form (\ref{dpp}).

The radiation polar part can be also reduced if we use scalar
constraint (\ref{wsk}) in two equivalent forms (motivated by (\ref{pp}))
 \be\label{wsk33}
  \sqrt{v}\left(\frac{r^3}{\sqrt{v}}\stackrel{\circ}{q}\! {^{AB}}{_{|| AB}}\right),{_3} +
 ({\dtwo}+2) \left[ {q}_3{^3} - \frac{\sqrt{v}}{2}\left(\frac{r}{\sqrt{v}}
 \stackrel{2}{q}\right),{_3} \right] =
 \sqrt{v}\left(\frac{r}{\sqrt{v}}\zeta \right),{_3} +{\cal B} \zeta  \ee
 \[
 \frac{1}{2}{\dtwo}({\dtwo}+2) \stackrel{2}{q} +
 2rv\left(r^2\stackrel{\circ}{q}\! {^{AB}}{_{|| AB}}\right),{_3} -
 {\dtwo}r^2\stackrel{\circ}{q}\! {^{AB}}{_{|| AB}} -
 2({\dtwo}+2) \left(r{q}^{3A}{_{|| A}}\right) = \]
 \be\label{wsk3A}
 = 2v \sqrt{v}\left(\frac{r}{\sqrt{v}}\zeta \right),{_3} 
 +2v{\cal B} \zeta -({\dtwo}+2){\cal B}^{-1} \zeta \ee
 Inserting (\ref{wsk33}), (\ref{wsk3A}) into (\ref{pp}) and
 integrating by parts we obtain
 \[ \mbox{radiation polar part}=
  \int_{V}{p}^3{_3} \left[ {q}_3{^3}
  + {\sqrt{v}}({\dtwo}+2)^{-1} \left(\frac{r^3}{\sqrt{v}}\stackrel{\circ}{q}\!
{^{AB}}{_{|| AB}}\right),{_3} 
  - \frac{\sqrt{v}}{2}\left(\frac{r}{\sqrt{v}}\stackrel{2}{q}\right),{_3} 
  \right] \; + \]
  \[
  + \int_{V} r{p}_{3A}{^{|| A}} \left[ \frac{1}{2}\stackrel{2}{q} +
 2r v{\dtwo} ^{-1} ({\dtwo}+2)^{-1} \left(r^2\stackrel{\circ}{q}\!
{^{AB}}{_{|| AB}}\right),{_3} -  ({\dtwo}+2)^{-1} \left(r^2\stackrel{\circ}{q}\!
{^{AB}}{_{|| AB}}\right) - 2  {\dtwo} ^{-1} \left(r{q}^{3A}{_{|| A}}\right) \right] + \]
\[ +
 \int_{\partial V} r{p}^3{_3}\left[ \frac{1}{2}\stackrel{2}{q} -
({\dtwo}+2)^{-1} \left(r^2\stackrel{\circ}{q}\! {^{AB}}{_{|| AB}}\right) \right]
-2 \int_{\partial V}  \left(r^2{p}^{3A}{_{|| A}}\right) {\dtwo} ^{-1}({\dtwo}+2)^{-1}
 \left(r^2\stackrel{\circ}{q}\! {^{AB}}{_{|| AB}}\right) = \]
\[ =
 \int_{\partial V} r{p}^3{_3} \frac{1}{2}\stackrel{2}{q}
- r\Pi {\dtwo} ^{-1}({\dtwo}+2)^{-1}
 \left(r^2\stackrel{\circ}{q}\! {^{AB}}{_{|| AB}}\right) 
+  \int_{V}{p}^3{_3}({\dtwo}+2)^{-1}\left[ 
 \sqrt{v}\left(\frac{r}{\sqrt{v}}\zeta \right),{_3} +{\cal B} \zeta  \right] +
 \] \[ +
   \int_{V} r{p}^{3A}{_{|| A}}{\dtwo} ^{-1} ({\dtwo}+2)^{-1}
    \left[  2\sqrt{v}\left(\frac{r}{\sqrt{v}}\zeta \right),{_3} +2{\cal B} \zeta
    - ({\dtwo}+2)v^{-1}{\cal B}^{-1} \zeta \right] = \]
\[
   =\int_{\partial V}r\left[{p}^3{_3} + 2{\dtwo} 
    ^{-1}\left(r{p}^{3A}{_{|| A}}\right)\right]
 ({\dtwo}+2)^{-1}\zeta + r{p}^3{_3} \frac{1}{2}\stackrel{2}{q}
- r\Pi {\dtwo} ^{-1}({\dtwo}+2)^{-1}
 \left(r^2\stackrel{\circ}{q}\! {^{AB}}{_{|| AB}}\right) +
 \]
\[
 + \int_{V} \left[-\frac{r}{\sqrt v}\left(\sqrt{v}\Pi\right)_{,3} +{\cal B}\Pi
-({\dtwo}+2){\cal B}^{-1}v^{-1} rp^{3A}{_{||A}} \right]
{\dtwo} ^{-1} ({\dtwo}+2)^{-1} \zeta = \]
\be\label{rpp}
=\int_{\partial V} r{p}^3{_3} \frac{1}{2}\stackrel{2}{q}
 +r\Pi {\dtwo} ^{-1} ({\dtwo}+2)^{-1}
 \left(\zeta - r^2\stackrel{\circ}{q}\! {^{AB}}{_{|| AB}}\right) 
 +\int_{V}[2 r^2\stackrel{\circ}{p}\!{^{AB}}{_{|| AB}} +{\cal B}\Pi]
{\dtwo} ^{-1} ({\dtwo}+2)^{-1} \zeta  \ee
where we used equality
\[ 2 r^2\stackrel{\circ}{p}\!{^{AB}}{_{|| AB}} +
 \frac{r}{\sqrt v}\left(\sqrt{v}\Pi\right)_{,3} + ({\dtwo}+2){\cal
B}^{-1}v^{-1} rp^{3A}{_{||A}}=0 \]
which is a simple consequence of the vector constraint (\ref{wwPi}),
(\ref{p333}) and (\ref{p3AA}).
It is easy to check that the volume terms in the final form of (\ref{rap})
and (\ref{rpp}) contain demanded gauge-invariant result.

Let us summarize the result we have proved
  \[
  \int_{V}{p}^{kl}{q}_{kl} = \mbox{monopole part} +
  \mbox{dipole part} + \mbox{radiation part} \]
  monopole part is given by (\ref{mpp}),
 the dipole part is a sum of (\ref{dap}) and (\ref{dpp}):
\[ \mbox{dipole part} = \mbox{dipole axial part}+ \mbox{dipole polar part} \]  
  and radiation part is a sum of (\ref{rap}) and (\ref{rpp}).
  
  \[   \mbox{radiation part in $V$}= \int_{V}
 \left[ 2 r^2 \stackrel{\circ}{p}\!{^{AB}}{_{|| AB}} +  {\cal B}\Pi \right]
\dtwo ^{-1} (\dtwo +2)^{-1} \zeta +  \]
\[  -2\int_{V}    (r^2 {p}^{3A|| B}\varepsilon_{AB}) \dtwo ^{-1}
\left[ {q}_{3A|| B}\varepsilon^{AB}  -
   (\dtwo +2)^{-1} ( r^2 \varepsilon^{AC}\stackrel{\circ}{q}\!
{_A{^B}}{_{|| BC}}),{_3} \right]  \]

  \[ \mbox{boundary terms}=
  \int_{\partial V} r v \Pi\dtwo ^{-1}\left(\dtwo +2-\frac{6m}r\right)^{-1}\! Q
 -r^2 p^{3A}{_{||A}}\dtwo ^{-1} \stackrel{2}{q} + 
  \] \[
 -2 \int_{\partial V}  (r^2{p}^{3A|| B}\varepsilon_{AB})
 {\dtwo} ^{-1}({\dtwo}+2)^{-1}
(r^2\varepsilon^{AC}\stackrel{\circ}{q}\! {_A{^B}}{_{|| BC}})  \]

\section{(2+1)-decomposition of the gauge for lapse and shift}

The gauge transformation (\ref{gauge4}) splits according to the
(2+1)-decomposition and we obtain the following gauge transformation law
which acts on lapse and shift:
\begin{eqnarray}
 \label{hxi00}
 h_{00}& \rightarrow & h_{00} +2\dot\xi_0 -\frac{2m}{r^2} \xi_{3}
\\ \label{hxi0A}
 h_{0A}& \rightarrow & h_{0A} + \xi_{0,A}+ \xi_{A,0} 
 \\ \label{hxi03}
 h_{03}& \rightarrow & h_{03} + \dot\xi_3- v\xi^0_{,3} 
\end{eqnarray}

\end{document}